%% file: main.tex
\documentclass[sigconf,9pt]{acmart} % Sensys requirement

\newif\ifACM
\ACMtrue

\input{0_packages}
\input{0_config}
\input{0_macros}
\begin{document}

\title{
Been There, Scanned That: Nostalgia-Driven LiDAR Compression for Self-Driving Cars
}
% Drive Through Memory: Nostalgia Meets LiDAR Compression for Self-Driving Cars}

% \title{Compression by Recall: Memory-Efficient LiDAR Data Representation for Autonomous Vehicles }

% Scanned Once, Perfect Forever: LiDAR Compression Reimagined

% }

\author[A. Khalid]{Ali Khalid}
\affiliation{%
\institution{Rochester Institute of Technology}
\country{}
}
\email{ak5013@rit.edu}

\author[J. Mobin]{Jaiaid Mobin}
% \authornote{Corresponding author}
\affiliation{
% \country{USA}
\institution{Rochester Institute of Technology}
\country{}
}
\email{jm5071@rit.edu}

\author[S. Appala]{Sumanth Rao Appala}
% \authornote{Corresponding author}
\affiliation{%
\institution{Vellore Institute of Technology}
\country{}
}
\email{sumanthrao.appala2021@vitstudent.ac.in}

\author[A. Maurya]{Avinash Maurya}
% \authornote{Corresponding author}
\affiliation{%
\institution{Rochester Institute of Technology}
\country{}
}
\email{am6429@rit.edu}

\author[S. Perez]{Stephany Berrio Perez}
% \authornote{Corresponding author}
\affiliation{%
\institution{The University of Sydney}
\country{}
}
\email{stephany.berrioperez@sydney.edu.au}

\author[M. Rafique]{M. Mustafa Rafique}
% \authornote{Corresponding author}
\affiliation{%
\institution{Rochester Institute of Technology}
\country{}
}
\email{mmrvcs@rit.edu}

\author[F. Ahmad]{Fawad Ahmad}
% \authornote{Corresponding author}
\affiliation{%
\institution{Rochester Institute of Technology}
\country{}
}
\email{fawad@cs.rit.edu}

\date{}

%% Uncomment the following line for camera-ready to add keywords,
%% acknowledgements etc.
\input{0_metadata}
\begin{abstract}
\input{abstract}
\end{abstract}

% CCS Concepts are already in 0_metadata.tex
%\keywords{Outdoor Pose Estimation, Augmented Reality, Aerial Mesh}

\maketitle
%%%%%%%%%%%%%%%%%%%%%%%%%%%%%%%%%% 80 CHAR %%%%%%%%%%%%%%%%%%%%%%%%%%%%%%%%%%%%%

%% Add as many lines here as you need
\input{1_intro}
\input{2_motivation}

\input{3_design_ali}
\input{4_evaluation}

\input{5_related}
\input{7_future_work}
\input{6_conclusion}
\input{8_Acknowledgment}

% \input{ToDo}

% \begin{acks}
% If needed
% \end{acks}

%{\Large\color{red}\bf 12 pages max., excluding references.}

% as many pages as necessary for bibliographic references
\newpage
\bibliographystyle{unsrt}
\bibliography{references.bib}

\end{document}

%% file: 0_packages.tex
\usepackage[T1]{fontenc}
\usepackage[utf8]{inputenc}
\usepackage{times}
\usepackage{upquote}

\usepackage{microtype}
\UseMicrotypeSet[protrusion]{basicmath} % disable protrusion for tt fonts
\usepackage{amsmath}
\usepackage{graphicx,grffile}
\usepackage{multirow}
\usepackage{xspace}
\usepackage{tabularx}
\usepackage{ragged2e}
\usepackage{booktabs}
\usepackage{paralist}
\usepackage[american]{babel}
\usepackage[shortlabels]{enumitem}
\usepackage{courier,color,wrapfig}
\usepackage{balance}
\usepackage{epstopdf}
\usepackage{url}
\usepackage{pifont}
\usepackage{tikz}
\usepackage{mathtools}
\usepackage[normalem]{ulem}
\usepackage{svg}
\usepackage{subcaption}

\usepackage{multirow}
\usepackage{tabularx}

\usepackage[linesnumbered,algoruled]{algorithm2e}
\usepackage{etoolbox}
\usepackage[textsize=small]{todonotes}
\usepackage[compact]{titlesec}
\usepackage{caption}
\usepackage[export]{adjustbox}
\usepackage[capitalize]{cleveref}
\crefformat{section}{§#2#1#3}
\crefname{enumi}{Step}{Steps}
\crefname{algorithm}{Alg.}{Algs.}

%% file: 0_config.tex
\makeatletter
\def\maxwidth{\ifdim\Gin@nat@width>\linewidth\linewidth\else\Gin@nat@width\fi}
\def\maxheight{\ifdim\Gin@nat@height>\textheight\textheight\else\Gin@nat@height\fi}
\makeatother
\setkeys{Gin}{width=\maxwidth,height=\maxheight,keepaspectratio}
\makeatletter
\g@addto@macro{\UrlBreaks}{\UrlOrds}
\makeatother

\makeatletter
\let\origsection\section
\let\origsubsection\subsection

\renewcommand\section{\@ifstar{\starsection}{\nostarsection}}
\renewcommand\subsection{\@ifstar{\starsubsection}{\nostarsubsection}}

\newcommand\sectionprelude{\vspace{1.5ex}}
\newcommand\sectionpostlude{\vspace{1.5ex}}
\newcommand\subsectionprelude{\vspace{0.75ex}}
\newcommand\subsectionpostlude{\vspace{0.75ex}}

\newcommand\nostarsection[1]{\sectionprelude\origsection{#1}\sectionpostlude}
\newcommand\starsection[1]{\sectionprelude\origsection*{#1}\sectionpostlude}

\newcommand\nostarsubsection[1]{\subsectionprelude\origsubsection{#1}\subsectionpostlude}
\newcommand\starsubsection[1]{\subsectionprelude\origsubsection*{#1}\subsectionpostlude}

\makeatother

\newcommand\paraspace{\vspace*{0.5ex}}
\providecommand\parab[1]{\paraspace\noindent\textbf{#1}}

%% file: 0_macros.tex
\newcommand{\sysname}{DejaView\xspace}

\newcommand{\scloud}{source cloud\xspace}
\newcommand{\tcloud}{reference cloud\xspace}
\newcommand{\sclouds}{source clouds\xspace}
\newcommand{\tclouds}{reference clouds\xspace}

\newcommand{\TCloud}{Reference Cloud\xspace}
\newcommand{\diff}{\texttt{diff}\xspace}
\newcommand{\wrt}{w.r.t\xspace}

\newcommand{\etc}{\emph{etc.}\xspace}
\newcommand{\ie}{\emph{i.e.,}\xspace}
\newcommand{\eg}{\emph{e.g.,}\xspace}

\newcommand{\secref}[1]{\S\ref{#1}}
\newcommand{\figref}[1]{Fig.~\ref{#1}}
\newcommand{\tabref}[1]{Tbl.~\ref{#1}}

\newcommand{\eqnref}[1]{Equation~\ref{#1}}

%% file: 0_metadata.tex
%%%%%%%%%%%%%%%%%%%%%%%%%%%%%%%%%%%%%%%%%%%%%%%%%%%%%%%%
%% ACMart stuff

\renewcommand\footnotetextcopyrightpermission[1]{} % removes footnote with conference info
\settopmatter{printacmref=false, printccs=false, printfolios=false}
% \settopmatter{printacmref=false}

\copyrightyear{2026}
\acmYear{2026}
\setcopyright{rightsretained}
\acmConference[Sensys '26]{ACM/IEEE Conference on Embedded Artificial Intelligence and Sensing Systems
}{May 11--14, 2026}{Saint-Malo, France}
\acmBooktitle{ACM/IEEE Conference on
Embedded Artificial Intelligence and Sensing Systems
, May 11--14, 2026, Saint-Malo, France}\acmDOI{}
\acmISBN{}

% \ifACM
% % \renewcommand\footnotetextcopyrightpermission[1]{} % removes footnote with conference info
% \setcopyright{none}
% % %\setcopyright{acmcopyright}
% % %\setcopyright{acmlicensed}
% % %\setcopyright{rightsretained}
% % %\setcopyright{usgov}
% % %\setcopyright{usgovmixed}
% % %\setcopyright{cagov}
% % %\setcopyright{cagovmixed}
% % \settopmatter{printacmref=false, printccs=false, printfolios=true}
% % % DOI
% \acmDOI{}
% % % ISBN
% \acmISBN{}
% % %Conference
% \acmConference[Submitted for review]{}
% \acmYear{2023}
% % \copyrightyear{}
% % %% {} with no args suppresses printing of the price
% \acmPrice{}
\pagestyle{plain}
% \fi

%%%%%%%%%%%%%%%%%%%%%%%%%%%%%%%%%%%%%%%%%%%%%%%%%%%%%%%%
% For accepted papers for ACM: change accordingly after uncommenting
\ifACM

% Use the code generated by the tool at http://dl.acm.org/ccs.cfm.

\ccsdesc[300]{Computer systems organization~Embedded and cyber-physical systems}
\ccsdesc[500]{Computing methodologies~Computer vision}
\ccsdesc[300]{Information systems~Information storage systems}

% \keywords{Autonomous Vehicles, LiDAR Data Compression, 3D Sensors}

\fi

%% file: abstract.tex
An autonomous vehicle can generate several terabytes of sensor data per day. 
A significant portion of this data consists of 3D point clouds produced by depth sensors such as LiDARs.  
This data must be transferred to cloud storage, where it is utilized for training machine learning models or conducting analyses, such as forensic investigations in the event of an accident.
To reduce network and storage costs, this paper introduces \sysname. 
Although prior work uses interframe redundancies to compress data, \sysname searches for and uses redundancies on larger temporal scales (days and months) for more effective compression. 
We designed \sysname with the insight that the operating area of autonomous vehicles is limited and that vehicles mostly traverse the same routes daily. 
Consequently, the 3D data they collect daily is likely similar to data they've captured in the past. 
To capture this, the core of \sysname is a \diff operation that compactly represents point clouds as delta \wrt 3D data from the past. 
% \ali{do you mean 'core to'?}
Using two months of LiDAR data, an end-to-end implementation of \sysname can compress point clouds by a factor of 210 at a reconstruction error of only 15 cm.
% significantly reducing the storage requirement for LiDAR data. 

% The key technical challenge in \sysname to ensure high compression without trading off reconstruction error and latency.
% For this, \sysname 
% The key technical challenge in \sysname is 
% To do this, \sysname stores point cloud data as a delta \wrt 

%% file: 1_intro.tex
\section{Introduction}
\vspace{-8pt}
Autonomous vehicles (AVs) rely on 3D sensors such as LiDARs, cameras, and Radars to understand their surroundings.
% themselves and perceive their surroundings using sensors such as cameras, LiDARs, and Radars \etc 
Beyond real-time operation, the data generated from these sensors is used offline for forensic analysis~\cite{feng2019autonomous, sharma2020towards, sharma2022cybersecurity, hoque2021avguard} to verify safety compliance and investigate insurance disputes. 
This data is also used to train and evaluate machine learning models for perception, prediction, and planning~\cite{mlusedata_muhammad2020deep,mlusedata_Fritsch2013ITSC,mlusedata_nuscenes,mlusedata_waymo}. 
As such, the sensor data generated from the AVs is \textit{uploaded offline, after the vehicle is parked, to cloud storage for long-term retention~\cite{cloudusedata_storage}}.
However, an AV's sensor suite can generate several terabytes (TBs) of data per day~\cite{wang2024quantitative,hoque2021avguard}. 
Among these sensors, a LiDAR, which generates 3D point clouds containing 3D points defined by their spatial coordinates and other attributes, is the most data-intensive sensor. 
% Of these sensors, a LiDAR which generates 3D point clouds contai
A 128-beam LiDAR produces over 2.6 million 3D points per second (equivalent to a data rate of 1~Gbps). 
With multiple LiDARs per vehicle (\eg 4 LiDARs on Waymo's 6th generation driver system~\cite{waymo6thgen}), per day, this amounts to 10's of terabytes of data.
% Of the sensor on-board an AV, with a conservative estimate, LiDARs generate over 8 TBs of data per day~\cite{wang2024quantitative}. 
% The output of a LiDAR is a 3D point cloud, a data structure consisting of a large number points defined by their 3D positions and other attributes like intensity and reflectivity.
% \fawad{Ali: Replace significant by some number or percentage?}
As the deployment of AVs expands, the sheer volume of sensor data will strain both network resources (for offline data transfer) and cloud infrastructure (for storage).
% For instance, with 4 LiDARs on-board Waymo vehicles~\cite{waymo6thgen}, this amounts of tens of terabytes per day, per vehicle.
% \stephany{For example, a 128-beam LiDAR can generate up to 1 Gbps of data when operating at 10Hz. 
% With multiple LiDARs mounted on AVs (\eg 4 LiDARs on Waymo vehicles~\cite{waymo6thgen}), per day, this amounts to tens of terabytes. }
% \fawad{mention lidar storage number. }
% This volume significantly stresses cloud storage capacity and wireless networks that transmit information from the vehicle to the cloud.

\begin{figure}[t]
    \centering
    \includegraphics[width=\columnwidth]{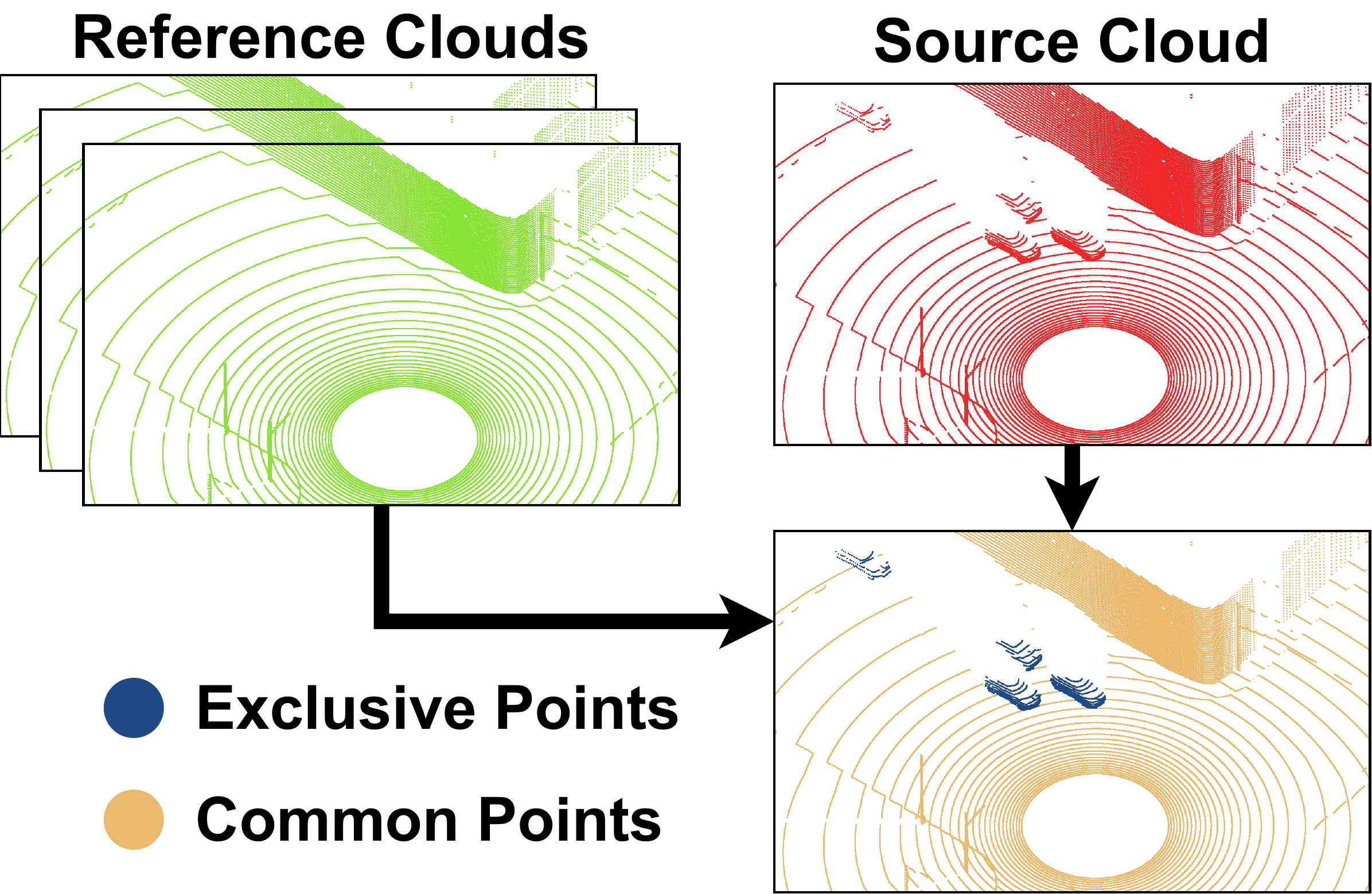} 
    \caption{The \scloud can be reconstructed using the common and exclusive points \wrt the \tcloud.}
    \label{fig:diff_concept}
\end{figure}

To reduce network transmission and cloud storage costs, in this paper, we build a framework for compressing 3D point clouds generated by LiDARs.
% LiDAR data.
% on compressing LiDAR data paper focuses on compressing LiDAR data to reduce network transmission and cloud storage costs. 
Once the vehicle is parked after its daily operation, this framework compresses 3D point clouds collected by the vehicle throughout the day and then transfers their compressed representations to the cloud for storage.
% transfer compressed LiDAR data from the vehicle to the cloud for permanent storage.
% , we compress LiDAR data at the end of a vehicle's trip. , then transfer and store the compressed representation at the cloud.
% the compressed representation to the cloud where it is stored  before the vehicle transfers it to the cloud for storage compress the data before the vehicle transfers it to the cloud for permanent storage.
Although there exist generic point-cloud compression techniques (\eg Draco~\cite{Draco}, Octree~\cite{octree}, and MPEG GPCC~\cite{mpeg_gpcc_software}), we posit that the characteristics of AV data present an opportunity for significantly higher compression ratios.
This opportunity stems from two observations: a) AVs, like human drivers, often traverse similar routes (\eg the daily commute to work), and b) AVs are mostly geofenced and their operation area is limited \cite{9484742}.
Consequently, \textit{ most of the data captured on any given day resembles data collected in the past days, weeks, or months, particularly for static environmental elements}, such as road structures, buildings, and traffic signs.
This redundancy across larger temporal scales presents a promising avenue to improve compression for AV generated LiDAR data.
% further compression of the data generated from AVs.

Existing compression techniques, such as those used in video streaming~\cite{video_compression_liu20242, video_compression_wiegand2003overview, video_hu2020improving, video_lu2019dvc}, typically focus on inter-frame compression over short timescales. 
These techniques store every subsequent frame as a delta (or difference) with respect to the previous frame.
Although effective in capturing frame-to-frame redundancies, these methods fail to take advantage of similarities across days, weeks, or months when vehicles operate in the same areas. 
Therefore, the overarching challenge in this paper is to identify redundancies across larger temporal scales and use them to compress 3D data.

This paper addresses the above challenge by using spatial hints in AVs.
Specifically, we make use of the fact that AVs are equipped with GPS and 3D maps, allowing them to position themselves precisely in the world.
Consequently, for any point cloud we want to compress (\figref{fig:diff_concept}: \scloud in red), we use the AV's position to retrieve the historically proximate point clouds (\figref{fig:diff_concept}: \tclouds in green). 
Then, we compare the \scloud with the \tclouds to identify two sets of points: a) \textit{common points} (\figref{fig:diff_concept}: brown points) that exist in both point clouds, and b) \textit{exclusive points} (\figref{fig:diff_concept}: blue points) that exist in \scloud but not in the \tclouds.
% point cloud but not the other.
% common and exclusive points (points that exist in one point cloud only) in each cloud.
In our proposed framework, we compactly store the \scloud using only the 3D positions of the \textit{exclusive points} and references (or pointers) to the \textit{common points}.
% The compressed version of the point cloud only contains the 3D positions of the \textit{exclusive points} and pointers (or references) to the \textit{common points}.
% In our representation, the compressed version of the \scloud contains a pointer to the \tcloud, indices of points exclusive to the \tcloud, and 3D positions of points exclusive to the \scloud.
This compressed version is sent over the network and stored in the cloud.
Although conceptually simple, implementing this presents significant challenges.

% \fawad{Ali: Do you think we can have a simple diagram that explains this concept?}
% \ali{Fawad: please review the above figure.}
% \fawad{Note to self: Give an intuitive explanation of the compressed format \ie references to common data and storage of new data; after that explain what this means}

\parab{Challenges.}
The core of our proposed solution is computing the difference between the \scloud and the \tcloud.
% \stephany{
This computation embodies a trade-off between compression ratio, reconstruction error, and latency.
% } 
Reconstruction error is the difference between an original point cloud and its reconstructed version after compression.

\underline{What to compute the difference against?} Compression performance depends on the number of exclusive points in the \scloud. Fewer exclusive points lead to higher compression ratios.
Compared to a single \tcloud, if we compute the difference against multiple consecutive \tclouds, the \scloud will share more common points with the \tclouds and hence fewer exclusive points.
However, comparing against multiple \tclouds can incur significant latency due to the increased number of point comparisons. 
Moreover, this inadvertently reduces compression ratios, as we show in \secref{s:motiv}.

% Moreover, this can inadvertently increase the exclusive points for the \tclouds, leading to lower compression ratios.

\underline{How to compute the difference? }To compute the difference between the point clouds, we use the nearest neighbor search to find, for each point in the \scloud, if an identical point exists in the \tcloud, and vice versa. 
For this, coarse-grained region-based approaches (\eg Octree~\cite{octree}) are fast, but could misclassify identical points if grouped into different regions. 
This leads to sub-optimal compression. 
Conversely, fine-grained point-wise approaches (\eg KD tree~\cite{kdtree_devillers2000geometric}) are more precise but incur significant latency.

% \stephany{A practical solution 
% must navigate the above tradeoffs to achieve high compression, low reconstruction error, and low latency.}

% \fawad{Ali and note to self: After adding the intuitive diagram and explanation about our compression representation, revisit the above challenges and change accordingly.}

\parab{Contributions.}
To address the above challenges, this paper makes the following contributions: 
\begin{itemize}[nosep,wide]

\item We propose a novel technique for compressing point clouds by leveraging redundancies in point clouds across temporal scales using their spatial relationships.

\item We demonstrate the efficacy of this approach by building an end-to-end system, \sysname, to compress AV LiDAR point clouds.

\item \sysname proposes a cascaded difference computation algorithm that uses a single \scloud and a collection of \tclouds to achieve high compression without trading off latency.

\item  \sysname proposed an accelerated nearest neighbor search algorithm that uses a coarse-grained fast search and a precise fine-grained search to enable fast compression without trading off reconstruction quality.

\end{itemize}

% \stephany {You need to prove this in the result section otherwise I'd refer to the results we have at the moment. Why two months?, you need to be careful of 'magic numbers' }
On a real-world vehicle LiDAR dataset collected over 2 months, consisting of 297K point clouds, an end-to-end implementation of \sysname achieves a compression ratio of 210 with a reconstruction error of less than 15~cm, significantly outperforming prior works.

% An end-to-end implementation of \sysname can compress real-world LiDAR data collected over $2$ months, consisting of $297K$ point clouds, with a compression ratio of $220$ and a reconstruction error of only $15~cm$, significantly better than prior work.

%% file: 2_motivation.tex
\section{Motivation and Background}
\label{s:motiv}

\begin{figure}[t]
    \centering
    \includegraphics[width=0.65\columnwidth]{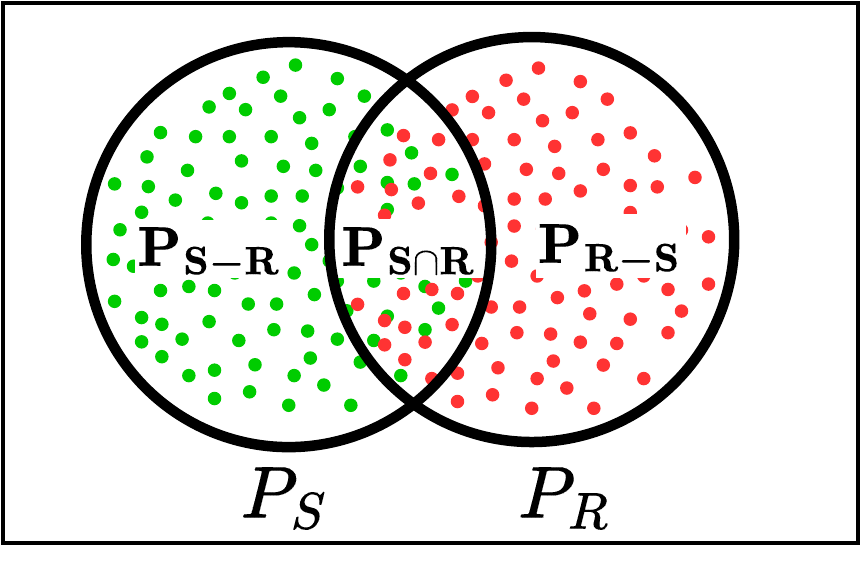} 
    \caption{Common and exclusive points in two point clouds.}
    \label{fig:diffVenn}
\end{figure}

% Background
    % Point cloud representation [x]
    % Background subtraction [x]
    % Compression using \diff
    % Strawman pipeline
        % Against a reference frame [single frame]
        % Against a 3D map [collection of frames]
    % Challenges

% \ali{Provide target application scenarios and more convincing motivations.}
% \fawad{Add missing references and then we are done.}
% \textcolor{blue}{

\parab{The Need for Storing Sensor Data. }
Beyond real-time perception, the sensor data generated by AVs is useful for a number of offline applications, including (but not limited to) forensic analysis, training, and verification of machine learning models.
For example, this data can be used to reconstruct scenarios leading up to events such as traffic accidents~\cite{shin2024recap}, unexpected vehicle behavior, corner cases, system failures, and even cyberattacks~\cite{sharma2020towards}. 
3D reconstructions of these events can enable comprehensive forensic analysis, resolve insurance disputes, and, most importantly, enhance root cause identification.
% }

% \textcolor{blue}{
% Beyond forensic applications, this data is critical for training, refining, and evaluating machine learning models across the autonomous driving stack \cite{mlusedata_muhammad2020deep,mlusedata_Fritsch2013ITSC,mlusedata_nuscenes,mlusedata_waymo} including perception, planning, and control.
Beyond forensic applications, large-scale real-world datasets collected daily from autonomous fleets are critical for training, refining, and evaluating machine learning models across the autonomous driving stack, including perception, planning, and control modules~\cite{mlusedata_muhammad2020deep,mlusedata_Fritsch2013ITSC,mlusedata_nuscenes,mlusedata_waymo}.
By integrating this data into simulation frameworks such as Waymo’s SimulationCity~\cite{waymo2021simulationcity}, developers can safely reproduce diverse traffic conditions, generate statistically representative edge cases, and continuously improve model robustness without requiring extensive on-road testing.
These real–world-grounded simulations allow ML models to adapt to evolving urban layouts, new mobility patterns, and rare driving events, ultimately accelerating the safe and scalable deployment of autonomous driving systems.
% } 
% Lastly, 
% For instance, e
% Consequently, the massive and continuous volume of multimodal sensor data makes storage and transmission resource-intensive. 
% Efficient compression is therefore essential for scalable and cost-effective data management without losing evidential or learning value.
% }

\parab{The Volume of AV LiDAR Data. } 
With multiple LiDARs mounted on a AV (e.g., four LiDARs on Waymo’s 6th-generation driver system~\cite{waymo6thgen}), just ten hours of operation, even with a very conservative estimate, can amount to 22~TBs of LiDAR data daily, 660 TB monthly, and 8 PB annually for a single vehicle. 
Using standard cloud storage pricing levels~\cite{awss3}, the monthly storage cost for a single vehicle's data would exceed \$10,000. For a fleet of just 100 vehicles, the annual storage costs would reach well over \$17 million.

Although these data are uploaded to cloud storage offline when the vehicle is parked, the sheer volume of data can lead to an unmanageable data transfer backlog. 
Even with a high-end WiFi 6 connection that achieves sustained speeds of 1.2 Gbps~\cite{wifi6}, uploading a single day of data would take nearly 40 hours—four times the amount of time to collect the data. 
Therefore, AV LiDAR data must be compressed to minimize both network transmission delays and long-term cloud storage cost.

\parab{3D Point Clouds. }
Depth perception sensors, such as LiDARs, generate collections of points in which each point is defined by its 3D position (x, y, z) and possibly other attributes such as color and intensity.
Point clouds are data-intensive and can be large in size.
A 128-beam LiDAR can produce more than 2.6 million 3D points per second, which is equivalent to a data rate of 1~Gbps. 

% For 22 TB/day per vehicle:

% Monthly data per vehicle = 22 TB × 30 days = 660 TB
% Yearly data per vehicle = 22 TB × 365 days = 8,030 TB ≈ 8 PB

% Using AWS S3 Standard Storage pricing tiers (approximate):

% First 50 TB/month: $0.023 per GB
% Next 450 TB/month: $0.022 per GB
% Over 500 TB/month: $0.021 per GB

% For one vehicle's monthly 660 TB:

% First 50 TB = 50,000 GB × $0.023 = $1,150
% Next 450 TB = 450,000 GB × $0.022 = $9,900
% Remaining 160 TB = 160,000 GB × $0.021 = $3,360
% Total monthly cost per vehicle ≈ $14,410

% \parab{The Need for Compressing AV LiDAR Data. }
% Quantify the storage and networks costs of one day's worth of AV LiDAR data
% Show that it is not practical to upload this data uncompressed as the number of vehicles increase
% One way to quantify: take some existing AV (e.g. Waymo) and determine daily data generated from there

\parab{Point Cloud Compression Techniques. }
Existing point cloud compression techniques like Octree~\cite{octree2buff} and Draco~\cite{Draco} leverage spatial redundancy within a point cloud to achieve compression. In octree-based methods, the point cloud is recursively subdivided into smaller volumetric units (voxels) to organize points hierarchically, which helps to reduce spatial redundancy. This octree representation helps to compress point clouds by enabling efficient encoding where only subdivisions with points are stored. 
Additionally, octree-based methods use interframe redundancies to achieve a better compression ratio. Draco, on the other hand, uses a KD tree to reorder points to put spatially close points together. This reordering helps Draco improve the efficiency of entropy encoding. Draco also uses quantization as a key technique to improve compression for point clouds. 

% \fawad{Ali, these descriptions are not that helpful. Can we rewrite this?}

\parab{Background Subtraction. }
Background subtraction, which we refer to as a \diff operation, finds the exclusive points in $P_{S}$ \wrt  $P_{R}$~\cite{octree2buff}. 
The operation \diff determines whether, for each point $s$ in $P_{S}$, there is an identical point $r$ in $P_{R}$.
To implement this, for each point $s$ in $P_{S}$, we find the nearest neighboring point $r$ in $P_{R}$ (\eqnref{eq:diff_equation}). 
If this point $r$ is within the distance range $d$ (we call this \textit{distance threshold}) from $s$, we consider $s$ and $r$ to be common or identical points ($P_{S \cap R}$ in \figref{fig:diffVenn}).
If not, then $s$ is a unique or exclusive point in $P_{S}$ \wrt $P_{R}$. 

\vspace{-8pt}

\begin{equation}
    \label{eq:diff_equation}
    \diff \left( P_{S}, P_{R} \right ) = \{ s \in P_{S} \mid \min_{r \in P_{R}} \| s - r \|_2 > d \}
\end{equation}

From the \diff operation, we have two sets of points: a) points common to $P_{S}$ and $P_{R}$ ($P_{S \cap R}$), and b) exclusive points in $P_{S}$ \wrt $P_{R}$ ($P_{S-R}$ in \figref{fig:diffVenn}).
If we wanted to find exclusive points in $P_{R}$ \wrt $P_{S}$ ($P_{R-S}$ in \figref{fig:diffVenn}), we would swap the positions of $P_{S}$ and $P_{R}$ in \eqnref{eq:diff_equation}. 
The \diff operation in this paper is used for point-cloud compression.

\parab{Point Cloud Compression using \diff. }
To compress the \scloud ($P_{S}$) using a \tcloud ($P_{R}$), 
we assume that $P_{R}$ is present for both compression and decompression.
In the first step, we perform a \diff operation of $P_{S}$ \wrt $P_{R}$ \ie \diff $\left ( P_{S}, P_{R} \right )$. 
From this operation, we obtain:
a) $P_{S-R}$, the exclusive points in $P_{S}$ \wrt $P_{R}$ and b) $P_{S \cap R}$, the common points between $P_{S}$ and $P_{R}$.
Next, we perform a reverse \diff operation \ie \diff $\left( P_{R}, P_{S} \right)$ to determine the exclusive points $P_{R-S}$ in $P_{R}$ \wrt $P_{S}$. 
Using only $P_{R}$ and the two sets of exclusive points ($P_{S-R}$ and $P_{R-S}$), we can reconstruct $P_{S}$ as shown below (\eqnref{eq:reconstruction}): 

 % \vspace{-4pt}

\begin{equation}
    \label{eq:reconstruction}
    P_{S} = P_{R} \: \text{-} \: P_{R-S}  \: \text{+} \:  P_{S-R} 
\end{equation}

Of the three sets of points that need to reconstruct $P_{S}$, we only need the 3D positions of the points exclusive to \scloud ($P_{S-R}$).
This is because \tcloud is already present at the end of the decompression and so are the points exclusive to it ($P_{R-S}$). 
To retrieve the exclusive points ($P_{R-S})$ from \tcloud, we only need their indices. 
Using the 3D positions of exclusive points in the \scloud ($P_{S-R}$) and indices of $P_{R-S}$, we can reconstruct $P_{S}$ (\eqnref{eq:reconstruction}).

% \fawad{Ali: Do we need a diagram to show this? MobiCom reviewers understood it.}
% \ali{I don't think so. }

\begin{figure}[t]
    \centering
    \includegraphics[width=\columnwidth]{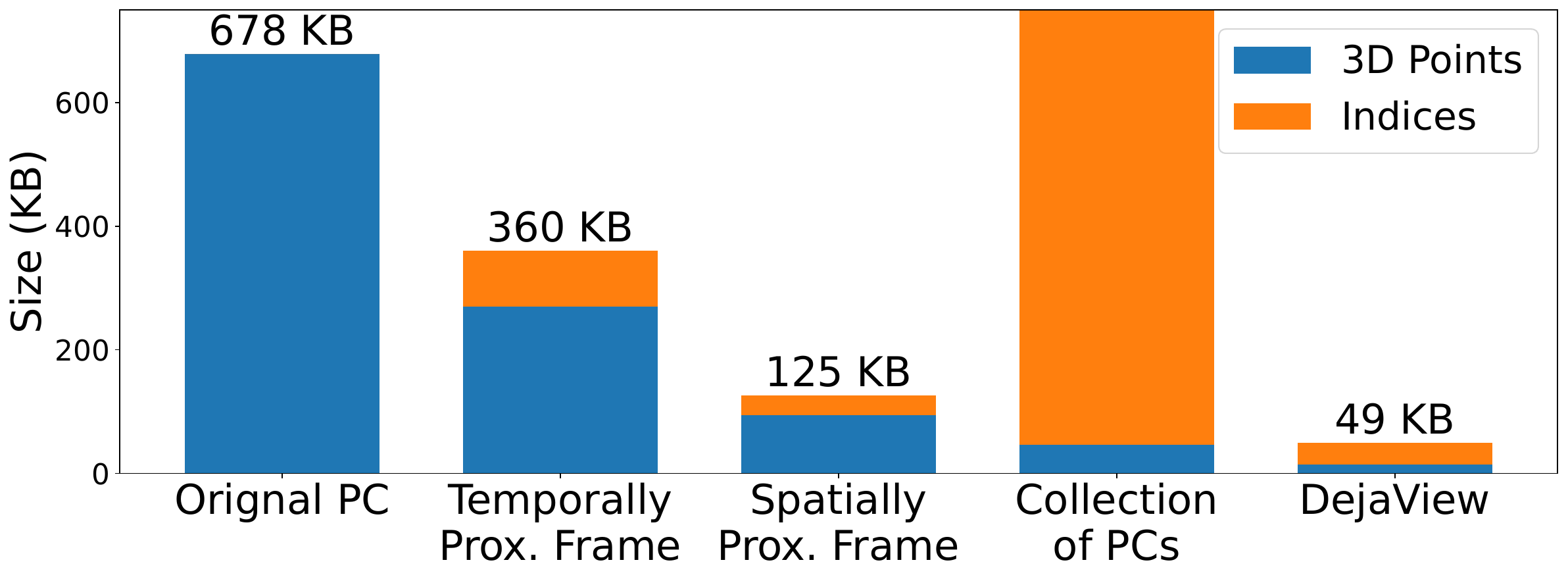} 
    \caption{Compression performance for strawman pipelines.}
    \label{fig:strawman_pipelines}
\end{figure}

\parab{Strawman Pipelines for Compression. }
% \stephany{I think you need to discuss what type of data you used? maybe some images would be good}
% \fawad{Ali: Stephany has a good point, could you please do that? Describe that you used a real-world dataset with x point clouds. Put it in red and let me know so that i can review.}
To motivate \sysname, we build three strawman pipelines that use the \diff operation to compress point clouds
and compare their performance (\figref{fig:strawman_pipelines}). 
We used 10 days of data generated from CARLA \cite{Dosovitskiy17} each with 500 point clouds to evaluate these pipelines. All pipelines use the same \textit{distance threshold} \ie 10~cm in this experiment.

\begin{figure*}[t]
    \centering
    \includegraphics[width=\textwidth]{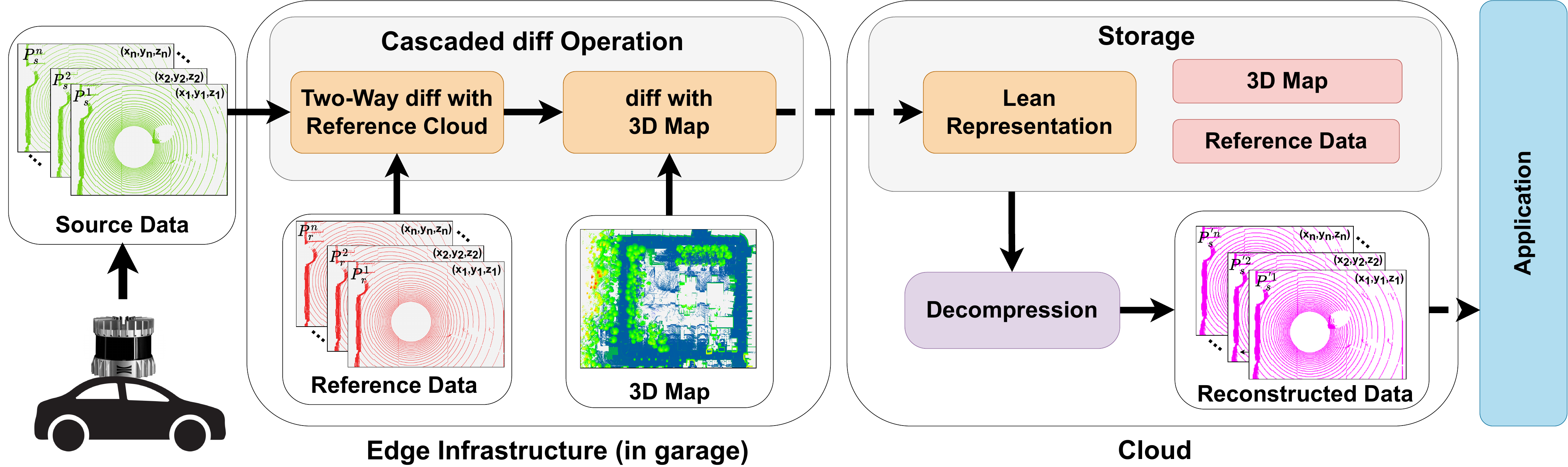} 
    \caption{Overview of \sysname. The AV generates source data, which is compressed at edge infrastructure (e.g., at garage or parking lot) to produce a lean representation. This lean representation is transmitted over the network to the cloud for storage. When requested by an application, the cloud decompresses the lean representation to reconstruct the data and delivers it to the application.}
    \label{fig:lidcom-overview}
    \vspace{-10pt}
\end{figure*}

\underline{Temporally Proximate Frame.} This approach adapts a technique similar to video streaming \ie representing \scloud as a difference \wrt to a point cloud from the previous frame.
Compared to the raw point cloud (first bar in \figref{fig:strawman_pipelines}), this approach (second bar) reduces the size of the data by 1.9x, on average. 
For each bar, the blue area represents the size of the 3D points exclusive to \scloud. 
On the other hand, orange represents the size of the indices for points that are exclusive to the \tcloud (in this case, the previous frame).
% whereas orange represents size of the indices. 
% \stephany{  I think it would be important here to explain what the indices are, what do they represent?}
% \fawad{Good point, reading this again, it feels like we have not properly explained \diff for compression yet.}
% \fawad{Instead of past and previous point cloud, can we do spatial and temporal proximate point clouds? Sounds more fancy ;)}

\underline{Spatially Proximate Frame. }
The second pipeline uses \sysname's idea \ie compress the \scloud using a spatially proximate \tcloud from the past.
This reduces the size of the point cloud by 5.4x (third bar in \figref{fig:strawman_pipelines}), on average. 
This is because \scloud shares more 3D points with a spatially proximate \tcloud (as opposed to a temporally proximate one). 
This has a two-fold effect. 
First, there are fewer exclusive points in the \scloud to store 3D positions for and fewer exclusive points in the \tcloud to store indices for.
% exc for and  points to store 3D positions for the  which leads to a higher compression ratio 
% As a result, there are fewer exclusive points to store. 
\sysname's \diff operation can reduce the point cloud size by 14x followed by a series of techniques (\secref{s:design}) to further increase the compression ratio to 210x.

\underline{A Collection of Point Clouds. } 
Intuitively, using a collection of \tclouds from the past should result in improved compression, because there is a higher probability of finding more common points and therefore fewer exclusive points for \scloud.
However, this is not true, as shown by the fourth bar in \figref{fig:strawman_pipelines}. 
In this experiment, we used 2000 \tclouds to find exclusive points in \scloud. 
Although this reduces the number of exclusive points in \scloud, it significantly increases the number of exclusive points to store for \tclouds. 
This is because now every \tcloud will have a separate set of exclusive points \wrt the \scloud.

As a result, the overhead of storing their indices is many times larger than the size of exclusive points of the \scloud.
Moreover, the size of the set of indices increases with the number of \tclouds.
So, \textit{this approach incurs four orders of magnitude higher latency} relative to comparing against a single-point cloud.

% \fawad{To understand this, I think we need that diagram I talked about in the introduction section.}

\parab{Challenges. }
\figref{fig:strawman_pipelines} shows the effectiveness of \sysname's proposed approach, achieving a 3.1x compression ratio. 
Although intuitively using more \tclouds should improve compression, our experiments show that this is not true and can be computationally expensive. 
The core challenge lies in finding the right balance: \sysname must identify more commonalities between the \scloud and \tclouds without introducing excessive overhead in managing these relationships.
Furthermore, the approach \sysname uses must find a balance between reconstruction accuracy and latency.

%% file: 3_design_ali.tex
% Overview Figures
%https://app.diagrams.net/#G1KolcjWPoWYmg1sZK7wRpgcpNSJl5ALp7#%7B%22pageId%22%3A%22j0OWW-qUYl0VbWe0Gmp2%22%7D

\section{\sysname Design}
\label{s:design}

\parab{Overview.}
\label{ss:overview}
% \ali{ Reviewer: What is the storage medium for on-vehicle raw point clouds, and can it meet the required write speed? Are the raw point clouds deleted after compression? }
% \textcolor{blue}{
To illustrate how \sysname operates, consider an AV that collects a sequence of point clouds (${P_{S}^1, P_{S}^2, \ldots, P_{S}^n}$) on a given day. The online perception module directly processes these point clouds for localization and scene understanding (object detection, tracking). 
The AV also temporarily stores these point clouds in high speed on-board memory before uploading them to the cloud for long-term storage. We refer to these point clouds as \sclouds (\figref{fig:lidcom-overview}).
% }
At the end of the operation of the AV, offline when the vehicle is parked in the garage, \sysname compresses \sclouds to build a compact representation for each \scloud. 
These compact representations of source clouds are sent over the network for cloud storage. 
When they are needed for other applications (\eg training, testing, or forensic analysis), \sysname decompressed them.
% At the cloud, they can be retrieved for other applications (\eg training, testing, or forensic analysis).

At its core, \sysname reduces the required network bandwidth and storage footprint by minimizing the number of points to store for a given \scloud. 
To do this, \sysname uses a reference dataset that contains \tclouds $\{P_{R}^1, P_{R}^2, \ldots, P_{R}^n\}$ (\secref{s:reference_dataset}). 
For each given \scloud, \sysname finds the approximate \tcloud from the reference dataset.
With a cascaded \diff operation (\secref{s:multi_stage_diff}), \sysname computes the difference between \scloud, its closest \tcloud, and the on-board 3D map\footnote{A dense 3D point cloud used by the AV for localization within their environment}. 
% \fawad{Not sure if we need to mention 3D map at this stage.}

\sysname reconstructs source point clouds on demand in the cloud in response to application requests and transmits them accordingly.
\sysname reconstructs \sclouds using their corresponding \tclouds in the reference dataset and the 3D map, as depicted in \figref{fig:lidcom-overview}.
We assume that the 3D map is available onboard the autonomous vehicle (AV), a common practice in the AV industry~\cite{waymo3Dmap,wevoloverreport}, as it is essential for navigation and operation. The reference dataset (\secref{s:reference_dataset}) is stored on edge infrastructure (e.g., in a garage where the AV is parked during compression) to reduce pressure on the vehicle’s onboard storage.
\sysname also assumes that both the 3D map and reference dataset are available in the cloud during decompression. To enable this, the AV uploads a copy of the 3D map and reference dataset to the cloud once during its lifetime. Maintaining these datasets on both the vehicle and the cloud reduces network load during both compression and decompression phases.

\subsection{Cascaded \diff Operation}
\label{s:multi_stage_diff}

%https://app.diagrams.net/#G1KolcjWPoWYmg1sZK7wRpgcpNSJl5ALp7#%7B%22pageId%22%3A%22eGFry3i63DbATwJBo8r-%22%7D

\begin{figure}[t]
    \centering
    \includegraphics[width=\columnwidth]{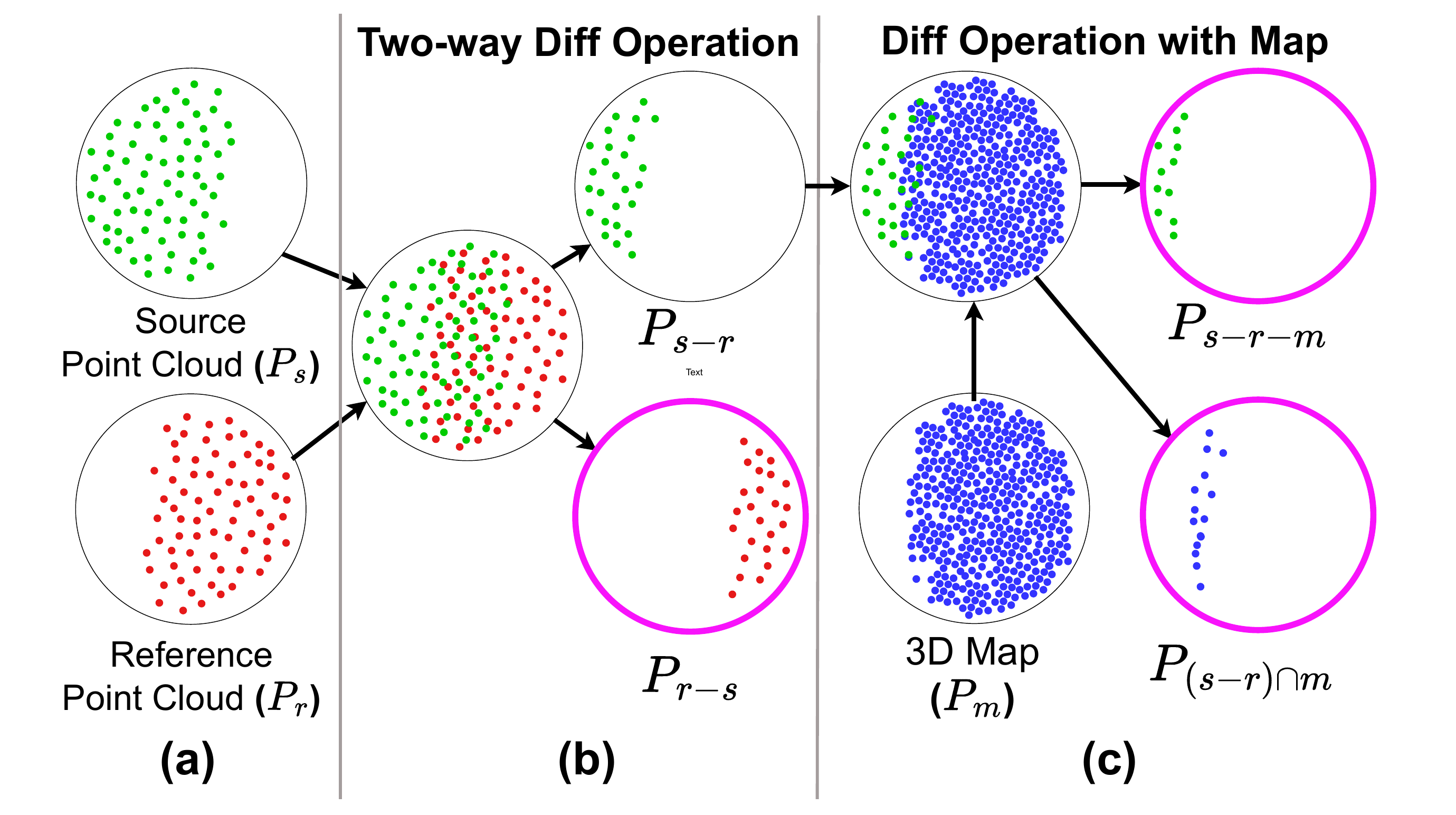} 
    \caption{\sysname's cascaded \diff operation uses a three step process to compute a compact representation for the \scloud using a \tcloud and a 3D map. The point clouds with purple outlines constitute the compact representation.}
    \label{fig:cascaded_diff}
\end{figure}

\sysname uses a cascaded \diff operation to compress \sclouds using \tclouds and a 3D map. 
This consists of: (a) a two-way operation \diff that compresses \scloud \wrt a \tcloud (\figref{fig:cascaded_diff}b), 
followed by (b) a single-way \diff operation that compresses the exclusive points of \scloud \wrt a 3D map (\figref{fig:cascaded_diff}c). 
In the second operation, \textit{instead of storing the exclusive points for the 3D map, \sysname stores the common points.}
This significantly reduces the exclusive points for \scloud without the additional overhead of the indices for \tclouds.
% \fawad{Might be too much detail here, can we save more of it for the sections below?}

\parab{\diff with a \TCloud. }
In the first stage of this operation, for a given \scloud ($P_{S}$), \sysname retrieves a \tcloud ($P_{R}$) from the reference dataset. 
The \tcloud is the closest point cloud in the 3D space to the \scloud (\secref{s:reference_dataset} 
discusses in more detail how we select a \tcloud). 

Then \sysname performs a two-way \diff operation between $P_{S}$ and $P_{R}$. 
The first \diff operation determines the exclusive points ($P_{S-R}$) in \scloud \wrt and \tcloud (\figref{fig:cascaded_diff}b).
The second \diff operation determines the exclusive points ($P_{R-S}$) in the \tcloud \wrt the \scloud (\figref{fig:cascaded_diff}b).
As a result of this two-way diff operation, the compact version of \scloud contains: a) 3D positions of exclusive points of \scloud ($P_{S-R}$), b) a pointer to \tcloud, and c) indices of the exclusive points of \tcloud ($P_{S-R}$). 
However, as demonstrated in \secref{s:motiv}, this operation only gives us a compression ratio of 5.4x. 
% \fawad{Not everyone might remember what Ps-r is, so would help to add the full-term instead.}

\begin{figure}[t]
    \centering
    \includegraphics[width=\columnwidth]{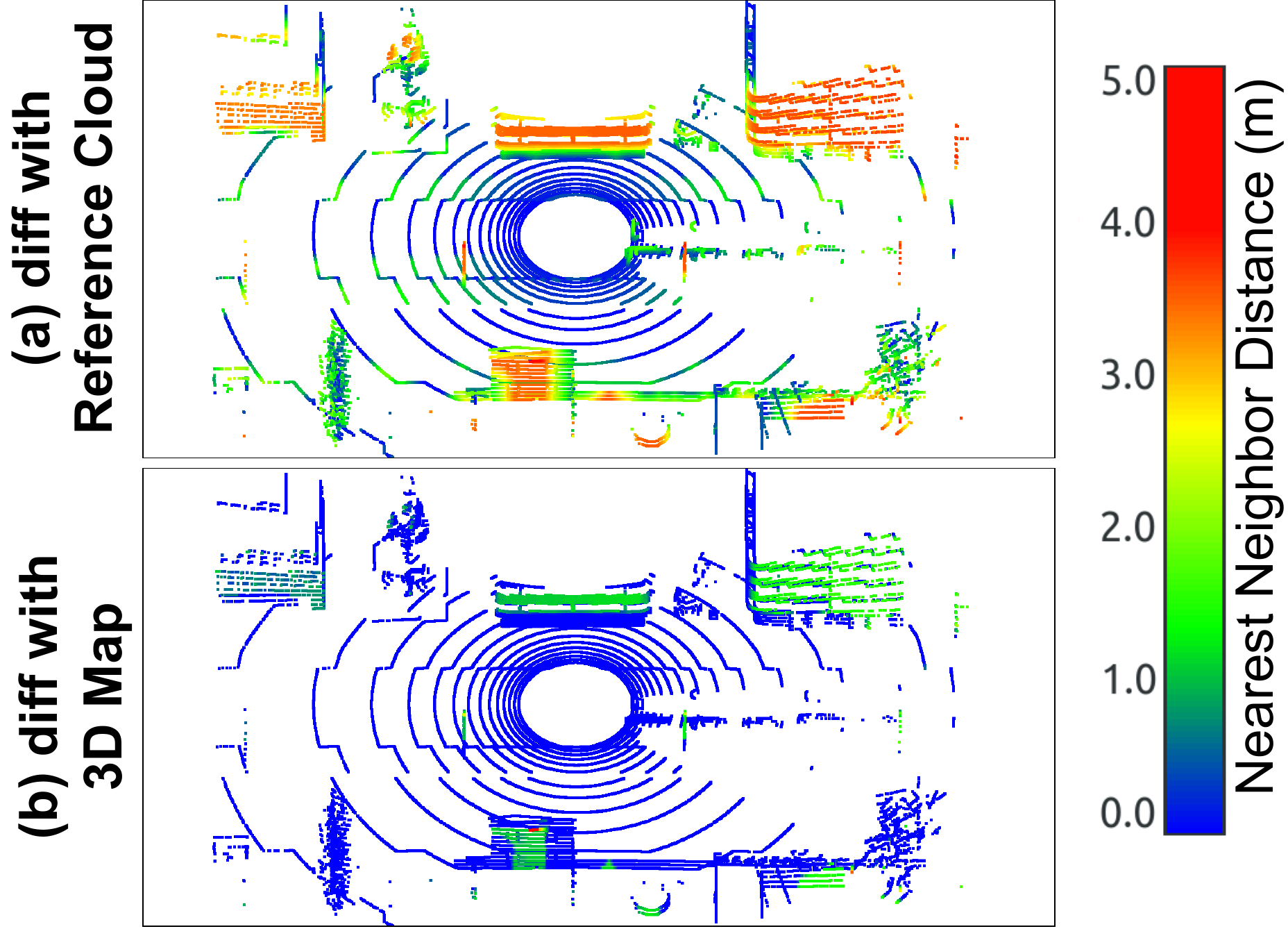} 
    \caption{\diff operation with a 3D map (b) can significantly reduce the number of points to store (non-blue points) as compared to a \diff against a \tcloud (a), which is effective only for points closer to the LiDAR sensor.}
    \label{fig:single_diff}
\end{figure}

The point density (points per volumetric unit) of LiDAR point clouds (\figref{fig:single_diff}) differs greatly throughout the point cloud. 
 Regions near the sensor are denser as compared to those further away because of the radial positioning of laser beams in a LiDAR's mechanical enclosure. 
\figref{fig:single_diff}a shows how this non-uniform distribution of points affects a \diff operation between \scloud and \tcloud \ie \diff $\left( P_{s}, P_{r} \right)$. 
% \kaleem{affects}
For a \scloud captured from a vehicle, we color-code every point to show the distance to its nearest neighbor in \tcloud. 
The blue points, in the denser regions near the sensor, are those for which \scloud finds an identical point in \tcloud. 
Points in regions further away (green, yellow, and red) are those for which we cannot find neighboring points close by, and hence are classified as exclusive points.

\parab{\diff with the 3D map. }
% \fawad{Good point, we wanted to do this evaluation but I belive we ran out of time.}
To improve compression, \sysname further reduces the number of exclusive points to store using a collection of point clouds or a 3D map ($P_{M}$). 
However, as discussed in \secref{s:motiv} and shown in \figref{fig:strawman_pipelines}, this is undesirable for two reasons. 
This operation can adversely affect the compression ratio because the set of exclusive points for the 3D map \wrt to the \scloud ($P_{M-S}$) can be significantly large.
Moreover, the second \diff operation (\ie \diff ($P_{M}$, $P_{S}$)) of the a two-way \diff operation between the 3D map and the source cloud , can incur significant latency because \sysname would need to find the nearest neighbors for all points in the 3D map. 
To address these challenges, \textit{\sysname uses an intelligently designed \diff operation that ensures both low latency and high compression ratio.} 

% However, this requires an intelligently designed \diff operation because  point cloud \ie a 3D map 
To optimize both the compression ratio and the latency, \sysname makes two careful design decisions. 
Firstly, \sysname uses only the set of exclusive points in \scloud ($P_{S-R}$) for the \diff operation.
Secondly, \sysname modifies the \diff operation to get the common points between the two input point clouds and uses them in the compact representation. 
% of  using two \diff operations, \sysname uses a single \diff operation.
Putting these together, \sysname performs only a one-way \diff operation using \scloud's exclusive points ($P_{S-R}$) against a vehicle's on-board 3D map ($P_{M}$) \ie \diff ($P_{S-R}$, $P_{M}$). 
This operation yields two sets of points (\figref{fig:single_diff} [c]): a) points exclusive to the \scloud \wrt both the \tcloud and the 3D map ($P_{S-R-M}$) and b) common points that exist in the \scloud and the 3D map but do not exist in the \tcloud ($P_{(S-R)\cap M}$). 
Using these sets of points, we can reconstruct $P_{S-R}$ (\eqnref{eq:map_recon}). 

\begin{equation}
    \label{eq:map_recon}
    P_{S-R} = P_{S-R-M} \: \text{+} \: P_{(R-S) \cap M} 
\end{equation}

This operation has multiple benefits. 
First, performing a \diff using exclusive points in the \scloud (\ie \diff ($P_{S-R}$, $P_{M}$)), instead of using the entire \scloud (\ie \diff ($P_{S}$, $P_{M}$)) significantly reduces latency. 
Second, because we can reconstruct $P_{S-R}$ using $P_{S-R-M}$ and $P_{(S-R)\cap M}$ (\eqnref{eq:map_recon}), we can avoid the second more compute-intensive \diff operation (\ie \diff ($P_{M}$, $P_{S-R}$)).
Third, it improves the compression ratio by replacing a large number of exclusive points in the map with common points between the 3D map and exclusive points in \scloud ($P_{(S-R)\cap M}$) in the compact representation.

The compact representation of \sysname of a \scloud consists of: a) 3D points exclusive to \scloud \wrt the \tcloud and a 3D map ($P_{S-R-M})$, b) reference to the \tcloud and indices of points exclusive to it \wrt to the \scloud ($P_{R-S}$), and c) indices of points in the 3D map that are common with exclusive points in \scloud \wrt to \tcloud $P_{(S-R) \cap M}$.

% https://drive.google.com/file/d/1KolcjWPoWYmg1sZK7wRpgcpNSJl5ALp7/view?usp=sharing

\parab{Further Compression. }
% \fawad{In the following, we should organize it so that we describe how to further compress points in separate section and indices in another.}
Although \sysname stores indices as opposed to 3D positions of exclusive points in \tcloud and common points with the 3D map, these sets of indices contain large integer values, primarily because of the large numbers of points in the point clouds.
This can adversely affect compression. 
To address this, \sysname uses delta encoding~\cite{suel2019delta} to convert large integer values to smaller values.
First, \sysname sorts the indices in descending order.
Then, it applies delta encoding, which stores every subsequent index as a function of the previous index (\eqnref{eq:delta_encoding}).

\begin{equation}
\label{eq:delta_encoding}
 DE \left(I\right) = \{i_1, \Delta i_2, \Delta i_3, ... , \Delta i_n \}  \\
\end{equation}
\[
where \quad \Delta i_x = i_{x-1} - i_x
\]

% \]

To improve compression, \sysname uses a hybrid technique to compress 3D points and indices.
\sysname uses Draco compression~\cite{Draco} for 3D points ($P_{S-R-M}$) and LZMA compression~\cite{lzma} for indices.
\sysname concatenates the two sets of delta-encoded indices ($P_{R-S}$ and $P_{(S-R) \cap M}$). 
It also appends their total counts to this list to separate the two at decompression time. 
On this list, it applies LZMA compression~\cite{lzma}.
After that, \sysname compresses 3D points using Draco and concatenates the LZMA and Draco outputs. It also prepends the Draco output size (in bytes) to the final output to separate Draco and LZMA parts during decoding. 
The compressed version of \scloud is sent over the network and stored in the cloud. 
In the following, we describe the decompression process when \scloud is needed for processing.

\parab{Decompression. }
To decompress, \sysname uses the first four bytes of the compressed \scloud to separate the Draco and LZMA data. 
Then, it decompresses the Draco output to retrieve the set of 3D points ($P_{S-R-M}$). 
Next, \sysname decompresses the LZMA stream. 
From this, it separates the two sets of indices ($P_{R-S}$ and $P_{(S-R) \cap M}$) using the first integer in the decompressed output. 
Then \sysname uses delta decoding to retrieve the original set of indices. 
Finally, it retrieves \tcloud and the 3D map and reconstructs \scloud (\eqnref{eq:recon_e2e}).

\begin{figure}[t]
    \centering
    \includegraphics[width=\columnwidth]{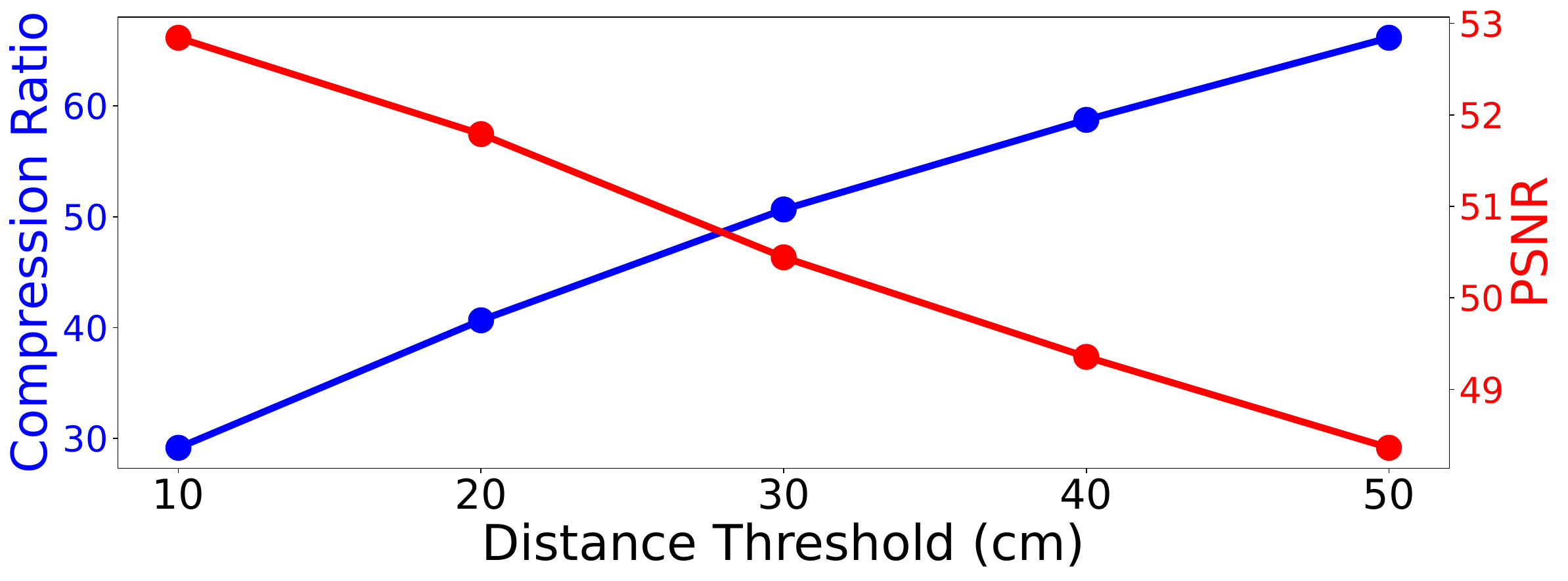} 
    \caption{As the distance threshold (d) increases, compression ratio increases at the cost of PSNR (\ie reconstruction error increases).}
    \label{fig:distance_threshold}
\end{figure}

% \begin{figure}[t]
%     \centering
%      % Subfigure 1
%     \begin{subfigure}{0.49\columnwidth}
%         \centering
%         \includegraphics[width=\textwidth]{figures/DistanceThresholdEffect_PSNR.pdf}
%         \caption{Compression ratio vs PSNR}
%         \label{fig:points_breakdown}
%     \end{subfigure}
%     % Subfigure 2
%     \begin{subfigure}{0.49\columnwidth}
%         \centering
%         \includegraphics[width=\textwidth]{figures/DistanceThresholdEffect_CD.pdf}
%         \caption{Compression ratio vs chamfer distance}
%         \label{fig:size_breakdown}
%     \end{subfigure}

%     \caption{As the distance threshold (d) increases, the compression ratio increases, but the reconstruction quality decreases (\ie lower PSNR and higher chamfer distance)}
%     \label{fig:distance_threshold}

% \end{figure}

% \begin{equation}
    % \label{eq:recon_first_step}
    % P_{s-r} = P_{(s-r) \cap m} +  P_{s-r-m} 
% \end{equation}?

\begin{equation}
    \label{eq:recon_e2e}
    P_{S} = P_{R} + P_{(S-R) \cap M} +  P_{S-R-M}  - P_{R-S} 
\end{equation}

% \ali{Reviewer: In addition, the system’s performance appears sensitive to the threshold used in the diff operation. Experimental results showing how performance changes as the threshold varies would strengthen the case for practical deployment.}
% \ali{I added CR \& CD vs distance threshold in the sensitivity analysis but i removed it later as this point is already discused here. Maybe we can move this paragraph to sensitivity analysis.}
% \fawad{Add a back reference in the eval section talking about what threshold does to compression ratio and PSNR.}
% \textcolor{blue}{
\parab{Compression Ratio Vs. Reconstruction Error. }
For points in \scloud and \tcloud to be identical, they must be within the \textit{distance threshold} $d$ from each other (\eqnref{eq:diff_equation}). 
This knob in \sysname controls the amount of compression we apply to \scloud. 
If the \textit{distance threshold} is high, more points in \scloud will be classified as common points and fewer will be exclusive points. 
Consequently, the compression ratio will increase (\figref{fig:distance_threshold}).
However, this comes at the cost of the Peak Signal-to-Noise Ratio (PSNR), a proxy that we use for the reconstruction error (more formally defined in \secref{s:evaluation}). 
Higher PSNR shows that the reconstructed \scloud is similar to the raw \scloud.
As the compression ratio increases because of the \textit{distance threshold}, the PSNR decreases.
% }

% \textcolor{blue}{
For example, the point $a$ in \scloud is at (0,0,0), whereas its nearest neighboring point $b$ in \tcloud is at (0,0,0.4).
If the \textit{distance threshold} is 0.5~m, then $a$ and $b$ would be classified as identical points. 
Thus, we would not need to store the 3D position of $a$, thus improving the compression ratio.
At reconstruction time, $a$ would be assigned the same position as $b$ \ie (0,0,0.4), which is 40~cm from its original position. 
As a result, the PSNR will be low.
Conversely, if the \textit{distance} threshold was smaller \ie 0.1~m, the $a$ would be an exclusive point. 
So, we would need to store the 3D position of $a$, thereby reducing the compression ratio. 
At reconstruction time, $a$ would be assigned its original position \ie (0,0,0), thus improving PSNR.
% }
% \fawad{It seems like this section is misplaced. But do not know where to place it instead.}

\subsection{Efficient \texttt{diff} Operation }

\parab{The Problem. }
To compress every \scloud, \sysname must perform three \diff operations, specifically, two with a \tcloud and one with a 3D map. 
These \diff operations must be fast
to guarantee same-day compression and upload. 
If not, this can cause a backlog in cloud transfer and storage (\secref{s:motiv}).

\parab{Point-based Techniques. }
Core to the \diff operation is the computation of the nearest neighboring point. 
In general, there are two techniques to determine the nearest neighbor.
Point-based approaches iterate through all points in one point cloud and determine their nearest neighboring point in the other cloud. 
These techniques use data structures, for example, the KD tree~\cite{kdtree_devillers2000geometric}, to speed up the nearest-neighbor search.

\parab{Region-based Techniques. }
%https://app.diagrams.net/#G1KolcjWPoWYmg1sZK7wRpgcpNSJl5ALp7#%7B%22pageId%22%3A%22MKvKtuANR8fbDIuldXqz%22%7D
\begin{figure}[t]
    \centering
    \includegraphics[width=\columnwidth]{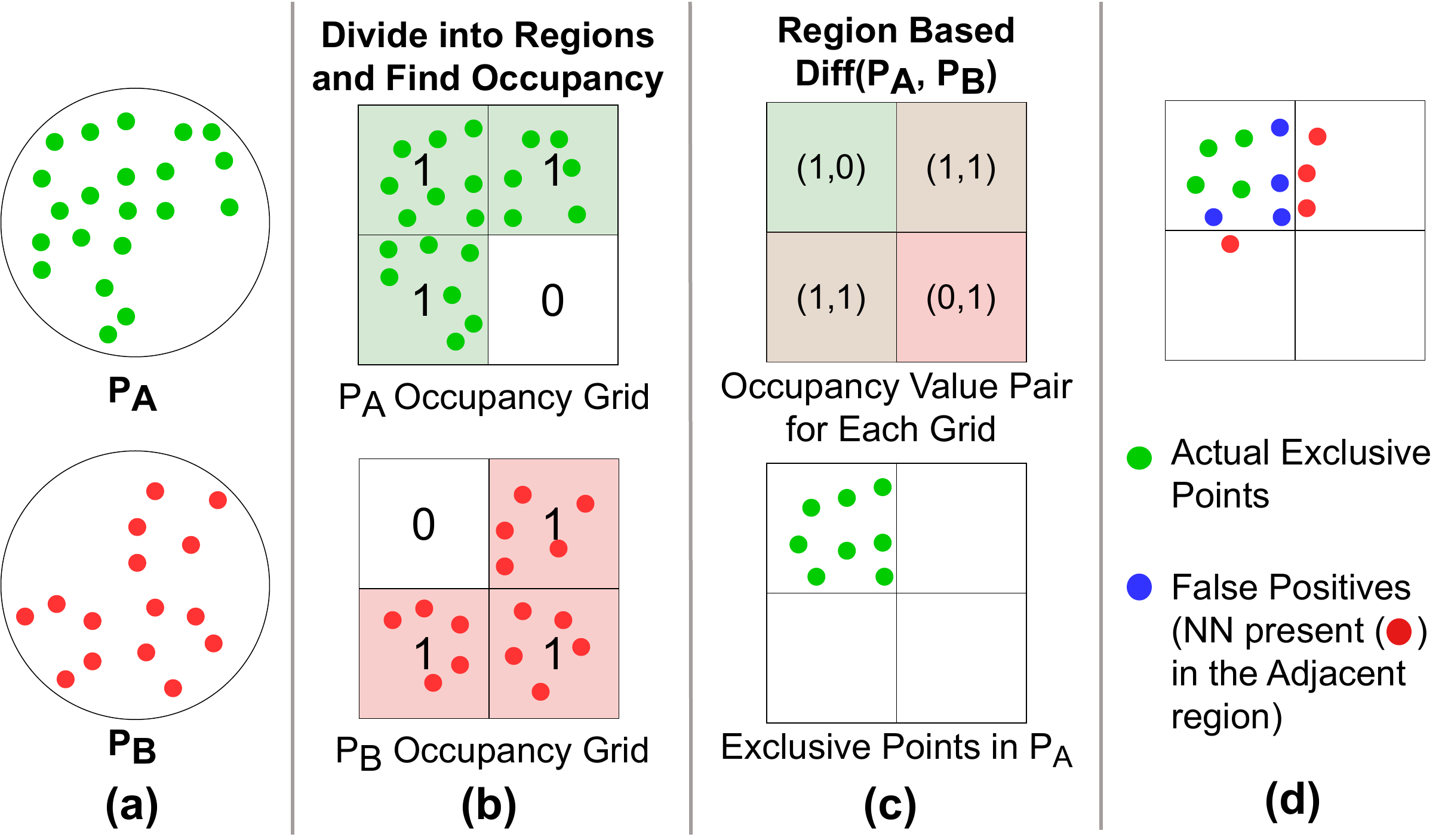} 
    \caption{\small Region-based \diff operations tend to aggressively classify exclusive points, often resulting in false positives.
    % \fawad{Ali: Change this to a single-way \diff \ie set 4th quadrant to 0,0 in (C)}
    }
    \label{fig:regionBasedDiff}
    \vspace{-5pt}
\end{figure}
Region-based techniques (Octree~\cite{octree2buff}, Voxel Grid~\cite{engel2004real, you2020pointwise})
partition point clouds into multiple regions (\figref{fig:regionBasedDiff}b). 
Each region has a flag that represents occupancy.
The flag is set if there are points within the region (\figref{fig:regionBasedDiff}b). 
% Using these, these techniques can perform comparisons at region-levels. 
If a region in the first point cloud is occupied but the corresponding region in the second point cloud is not, then all points in the first point cloud's region are classified as exclusive. 
% In \figref{fig:regionBasedDiff}c, the top-right and bottom-left regions . 
As such, in \figref{fig:regionBasedDiff}c, the top-left region of $P_{A}$ is classified as exclusive \wrt $P_{B}$, and the bottom-right region of $P_{B}$ is classified as exclusive \wrt $P_{A}$. 
However, a point's nearest neighbor does not necessarily have to be in the same region. 
For example, multiple points in the top-left region of $P_{A}$ have nearest neighbors in the top-right region of $P_{B}$. 
Because region-based techniques do not consider this, these points are misclassified as exclusive points (blue points in \figref{fig:regionBasedDiff}d).

%https://app.diagrams.net/#G1KolcjWPoWYmg1sZK7wRpgcpNSJl5ALp7#%7B%22pageId%22%3A%22MKvKtuANR8fbDIuldXqz%22%7D
\begin{figure}[t]
    \centering
    \includegraphics[width=\columnwidth]{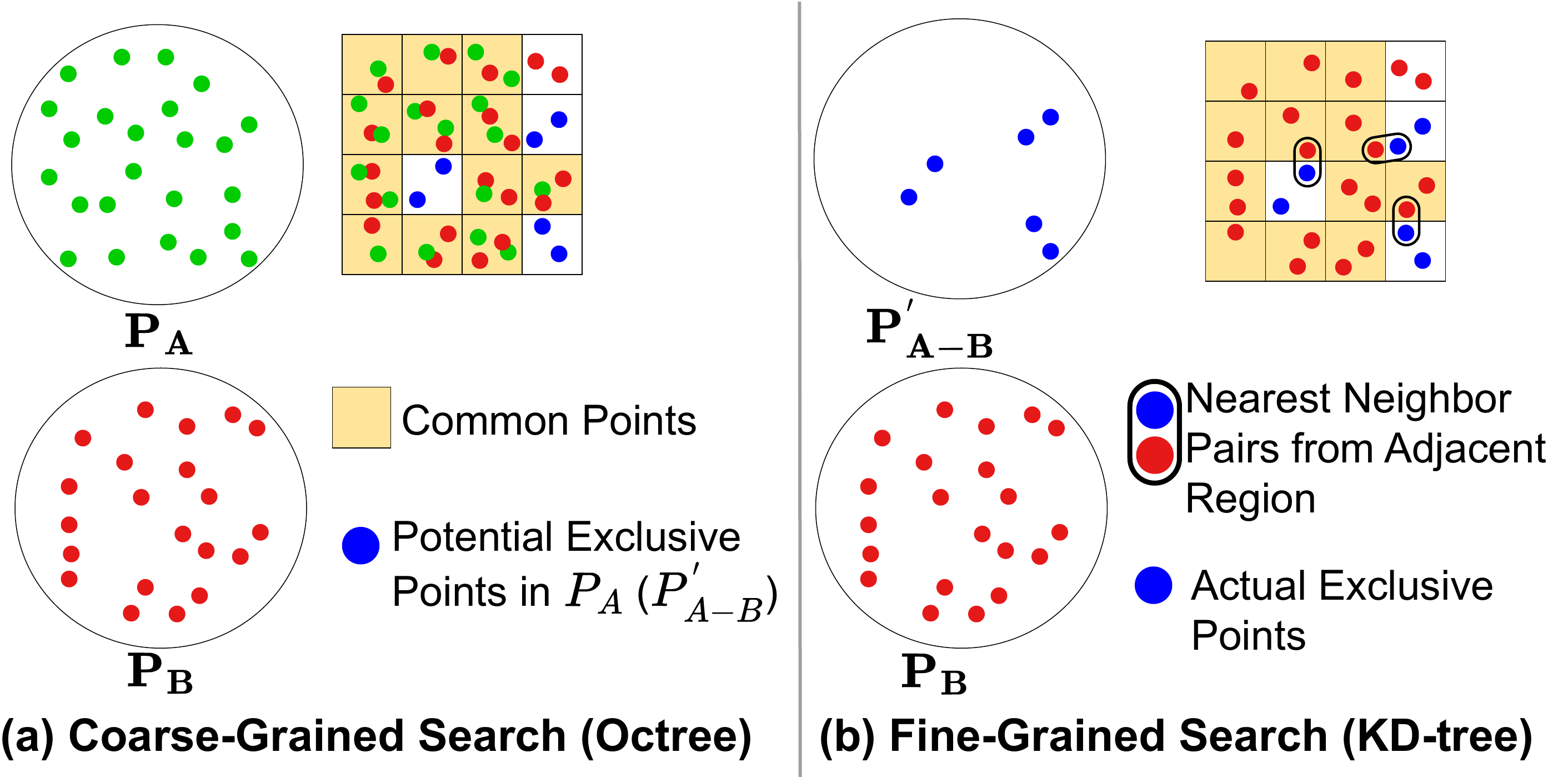} 
    \caption{\sysname uses a hybrid technique that uses coarse-grained and fine-grained searches to ensure low latency and high compression.}
    \label{fig:hybrid_diff}
    % \vspace{-10pt}
\end{figure}

\begin{table}[t]
    \centering
    \footnotesize
     \resizebox{\columnwidth}{!}{%
    \begin{tabular}{|c|c|c|c|}
        \hline
        \diff Technique & Latency (ms) & Compression Ratio\\ \hline \hline
        Point-based (KD-tree~\cite{kdtree_devillers2000geometric}) & 663 & 3.4 \\ \hline
        Region-based (Octree~\cite{octree2buff}) &  145 & 2.3 \\ \hline
        \sysname  & 405 & 3.4\\ \hline
    \end{tabular}}
        \caption{Point-based approaches can do better compression (at the cost of latency) whereas region-based techniques are faster (at the cost of compression). \sysname's hybrid approach is both fast and can do better compression.
        % \fawad{Ali: Revisit this table.}
        }

    \label{tab:efficient_diff}
\end{table}

\parab{Quantitative Comparison. }
To compare the two approaches, we used them to perform a two-way operation \diff between multiple \sclouds and \tclouds. 
For this experiment, we measured the compute latency and compression ratio for these operations (\tabref{tab:efficient_diff}). 
Point-based techniques have higher compression ratios but can be slow. 
Region-based techniques are faster but have lower compression because they tend to have false positive exclusive points (as explained above). 
As a result, region-based techniques overestimate the number of exclusive points, leading to a decrease in compression.
% (\tabref{tab:efficient_diff}). 

\parab{Our Approach. }
To address these challenges, \sysname uses a hybrid search in its \diff operations. 
This consists of a coarse-grained search followed by a fine-grained search. 
Assume \sysname performs a \diff operation between $P_{A}$ and $P_{B}$. 
To do this, \sysname loads both point clouds into an octree and performs a region-based search. 
This operation quickly identifies the \textit{ potential} exclusive points ($P_{A-B}^{'})$ between two points clouds (\figref{fig:hybrid_diff}a). 
Knowing that this can produce false positive results, \sysname refines this with a fine-grained search. 
For this, \sysname loads $P_{B}$ onto a KD tree. 
Then, for each \textit{potentially} exclusive point ($P_{A-B}^{'})$, it queries the KD tree for its nearest neighbor in $P_{B}$ (\figref{fig:hybrid_diff}b). 
Once it retrieves the nearest neighbor of the point, it classifies it as an \textit{actual} exclusive point if it does not lie within the predefined \textit{distance threshold}. 
Otherwise, it is classified as an identical or common point.
In this way, \sysname can remove false exclusive points.

\sysname's hybrid \diff operation is both fast and ensures high compression (\tabref{tab:efficient_diff}). 
The coarse-grained search ensures high computational efficiency. 
The fine-grained search mitigates their propensity to overestimate exclusive points. 
Moreover, since the fine-grained search is used for fewer points, it reduces the overall latency overhead.
Populating a KD tree can be computationally expensive. 
Because all \sclouds perform a \diff against the same 3D map, \sysname preloads the 3D map into a KD-tree representation off-line and reuses this for every frame.

\label{s:reference_dataset}

\subsection{Reference Dataset}

\sysname uses a reference dataset consisting of \tclouds $\{P_{R}^1, P_{R}^2, \ldots, P_{R}^n\}$ to compress and decompress \sclouds. 
This data set consists of point clouds that AVs have collected in the past for a given region on any given day. 
We assume that this data set is stored in uncompressed format in the edge compute (\eg the parking garage) and in the cloud. 
AVs within a similar geographic region (\eg town or city) can share this data set. 
Each \tcloud has a pose (3D position and rotation) that describes where the point cloud was captured.
% We assume this dataset is available at an edge server in the garage where the AV is parked, and at the cloud.

Selecting \tcloud for a given \scloud is crucial to maximize the compression ratio. 
Ideally, if the \tcloud points are in the same 3D positions as those in \scloud, this would yield maximum compression. 
However, determining point-level similarity can be computationally expensive. 
Instead, \sysname uses spatial proximity as a proxy for point-level similarity. 
This is based on the intuition that the static parts of the environment will be structurally similar across larger temporal scales.

\begin{figure}[t]
    \centering
    \includegraphics[width=\columnwidth]{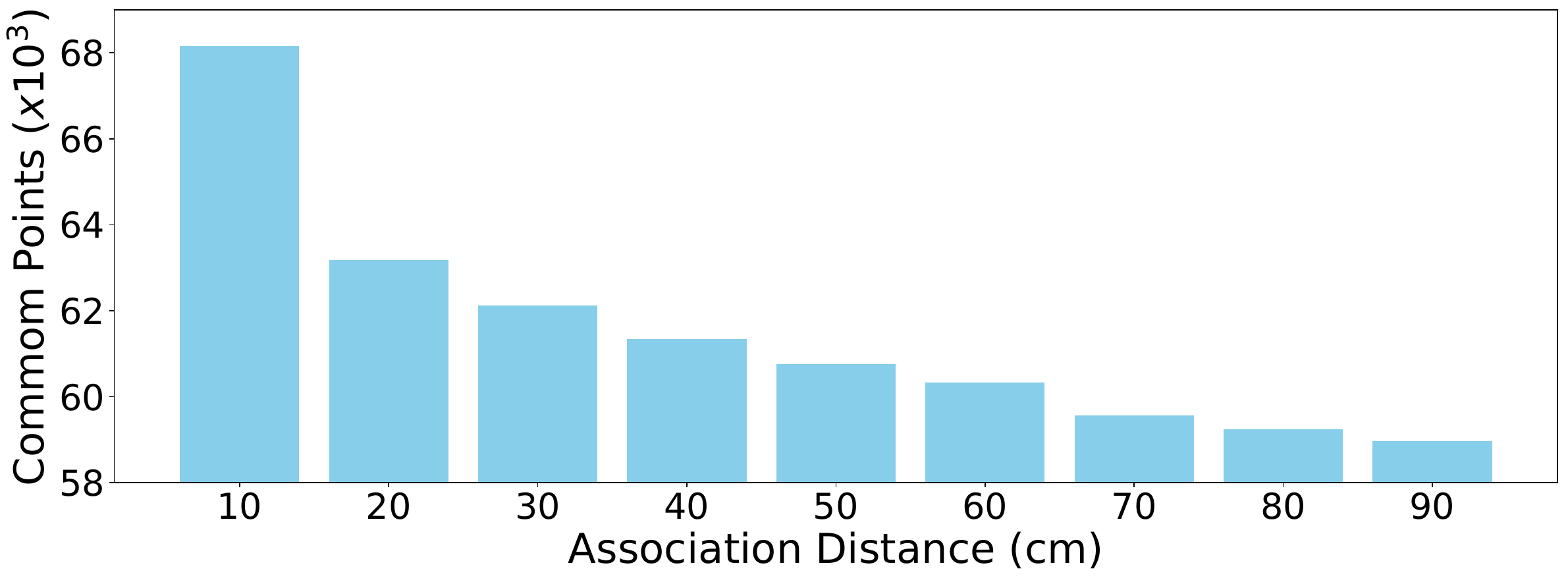} 
    \caption{Close-by \scloud and \tcloud pairs (lower association distances) tend to have more common points and fewer exclusive points, hence leading to better compression.}
    \label{fig:associationDistdEffect}
\end{figure}

For quantitative evaluation, we used 2000 pairs of \scloud and \tcloud and measured the number of common points by changing the 3D distance (association distance) between them. 
As \figref{fig:associationDistdEffect} shows, if \scloud and \tcloud are close together, they have a larger number of common points between them, leading to higher compression ratios. 
\figref{fig:associationDistdEffect} also shows a sharp decrease in common points if the distance between \scloud and \tcloud is greater than 10~cm.

These 3D positions for \scloud and \tcloud were already computed when the AVs collected these point clouds during their normal operation. 
To reduce latency, \sysname performs this association using vectorized calculations. 
The 3D positions of \sclouds and \tclouds are loaded in separate matrices. 
Then it computes the Euclidean distances for all pairs of \scloud and \tcloud by matrix multiplication. 
The \tcloud with the lowest distance from \scloud is associated with it.

%% file: 4_evaluation.tex
\section{Evaluation}
\label{s:evaluation}

\subsection{Methodology}
\label{ss:eval_methodology}

\parab{Implementation. }
We have implemented \sysname as three main modules using C++ and Python. 
The compression and decompression modules are written in C++.
These use the Point Cloud Library (PCL~\cite{Rusu_ICRA2011_PCL}) for point cloud operations.
The third module, which associates \sclouds with \tclouds, is a Python script.
For compression, we use Draco~\cite{Draco} and LZMA~\cite{lzma}. 
To build the 3D map on board, we use Fast-LIO2~\cite{fast-lio2}, a LiDAR SLAM algorithm. 
Then, we localize the point clouds in this 3D map using a normal distribution transform (NDT)~\cite{biber2003normal}.
We embed our compression and decompression modules for higher throughput in OpenMP library's wrappers.
Based on the number of cores available on a machine, these can compress multiple \sclouds concurrently.
\sysname's code and dataset are open-source\footnote{GitHub Repository: https://github.com/nsslofficial/DejaView}.
% \stephany{Would this code be available? if so, we should have the github repo (even it's only the readme by now)}

\begin{figure}[t]
    % \vspace{-2pt}
    \centering
    \includegraphics[width=\columnwidth]{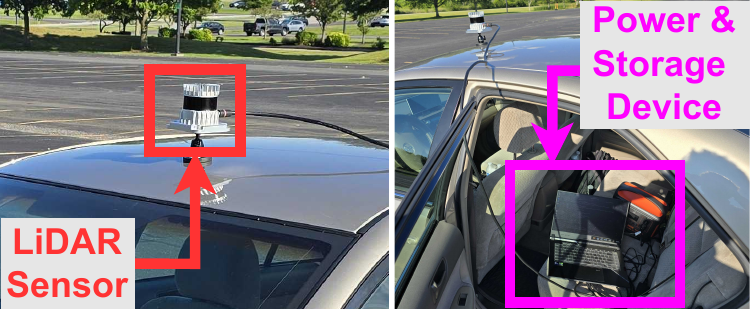} 
    \caption{Real-world data collection setup}
    \label{fig:car_data_collection}
\end{figure}

\parab{Real-world Dataset.} 
To evaluate \sysname, we collected our own data set by mounting an Ouster OS1-128 beam LiDAR~\cite{ouster_os1_lidar_sensor} on top of a vehicle (\figref{fig:car_data_collection}). 
We drove daily on the same route around our campus
for two months, accumulating over 297,000 point clouds. 
Using data for a single day from this dataset, we built a 3D map using Fast-LIO2~\cite{fast-lio2}.

\parab{Synthetic Traces.}
We complemented our real-world dataset with synthetic LiDAR traces collected from CARLA~\cite{Dosovitskiy17}, a photorealistic autonomous driving simulator. 
CARLA provides us with accurate ground truth for multiple applications that we do not have access to in the real world. 
For example, CARLA gives us accurate ground truth for localization, 3D object detection, and segmentation, which we use to evaluate \sysname.
Moreover, CARLA is flexible because it allows us to perform a sensitivity analysis for \sysname by varying the number of channels and noise in a LiDAR and varying the traffic density.
% channels and noise in a LiDAR and.
To match our real-world data collection, we collected 61 LiDAR traces along the same route, a total of more than 171,000 point clouds.

\parab{Evaluation Platform. }
We evaluated \sysname using a desktop with a 16-core Intel Xeon Silver 4114 CPU, 32 GB RAM, and Quadro P1000 GPU. 
This compute platform performs the compression and is connected to the internet through a wired network.
We send the compressed point clouds to a CloudLab~\cite{cloudlab}
server approximately 2000 miles away.

% \begin{figure}[t]
%     \centering
%     % Subfigure 1
%     \begin{subfigure}{0.49\columnwidth}
%         \centering
%         \includegraphics[width=\textwidth]{figures/end-to-end_real_CD_1.pdf}
%         \caption{\sysname has higher compression ratios at a given chamfer distances relative to other methods}
%         \label{fig:real_CD}
%     \end{subfigure}
%     % Subfigure 2
%     \begin{subfigure}{0.49\columnwidth}
%         \centering
%         \includegraphics[width=\textwidth]{figures/end-to-end_real_PSNR_1.pdf}
%         \caption{ \sysname achieves a higher compression ratio at a given PSNR as compared to Octree and Draco.}
%         \label{fig:real_PSNR}
%     \end{subfigure}
   
%     \caption{ Compression ratio vs. reconstruction quality for real-world dataset.}
% \end{figure}

% \begin{figure}[t]
%     \centering
%     \includegraphics[width=0.7\columnwidth]{figures/end-to-end_real_CD.pdf} 
%     \caption{On real-world datasets, \sysname has higher compression ratios at lower Chamfer distances relative to Draco and Octree.}
%     \label{fig:real_CD}
% \end{figure}

% \begin{figure}[t]
%     \centering
%     \includegraphics[width=0.65\columnwidth]{figures/end-to-end_real_PSNR.pdf} 
%     \caption{On real-world datasets, \sysname achieves higher compression ratio at lower PSNR as compared to Octree and Draco.}
%     \label{fig:real_PSNR}
% \end{figure}

\parab{Evaluation Metrics.} 
We use the following metrics to evaluate the compression and reconstruction error of \sysname.

\underline{Compression Ratio.} The compression ratio is the ratio of the original and compressed sizes \sclouds. Higher compression ratios are better.

\underline{Chamfer Distance. } We use the symmetric Chamfer distance \cite{wu2021balanced} to measure \sysname's reconstruction error.
It is the least-mean-square-distance between every point in one point cloud to its nearest neighboring point in the other point cloud.
More formally, given an original point cloud \(P\) and a
reconstructed point cloud \(P^{'}\), the symmetric Chamfer distance is defined in Equation \ref{eq:sym-chamfer-distance}. 
A Chamfer distance of 0 indicates perfect reconstruction.

\begin{equation}
    \label{eq:chamfer-distance}
    \mathbf{CD}(P, P^{'}) = \frac{1}{\left| P \right|} \sum_{i}^{} \min_{j} \|p_{i}-p^{'}_{j} \|_{2}
\end{equation}

\begin{equation}
    \label{eq:sym-chamfer-distance}
    \mathbf{CD_{sym}}(P, P^{'}) =  \{\mathbf{CD}(P, P^{'}) + \mathbf{CD}(P^{'}, P) \}  / 2
\end{equation}

\underline{Point-to-Plane Peak Signal-to-Noise Ratio (PSNR). } 
The symmetric point-to-plane peak signal-to-noise ratio (which we refer to as PSNR)~\cite{tian2017geometric} measures how well the points of one cloud align with the surfaces of the other cloud. 
Unlike the Chamfer distance, which measures the raw positions of 3D points, PSNR measures the retention of surface geometry information in the reconstructed point cloud.

\begin{figure*}[t]
    \centering
    \includegraphics[width=\textwidth]{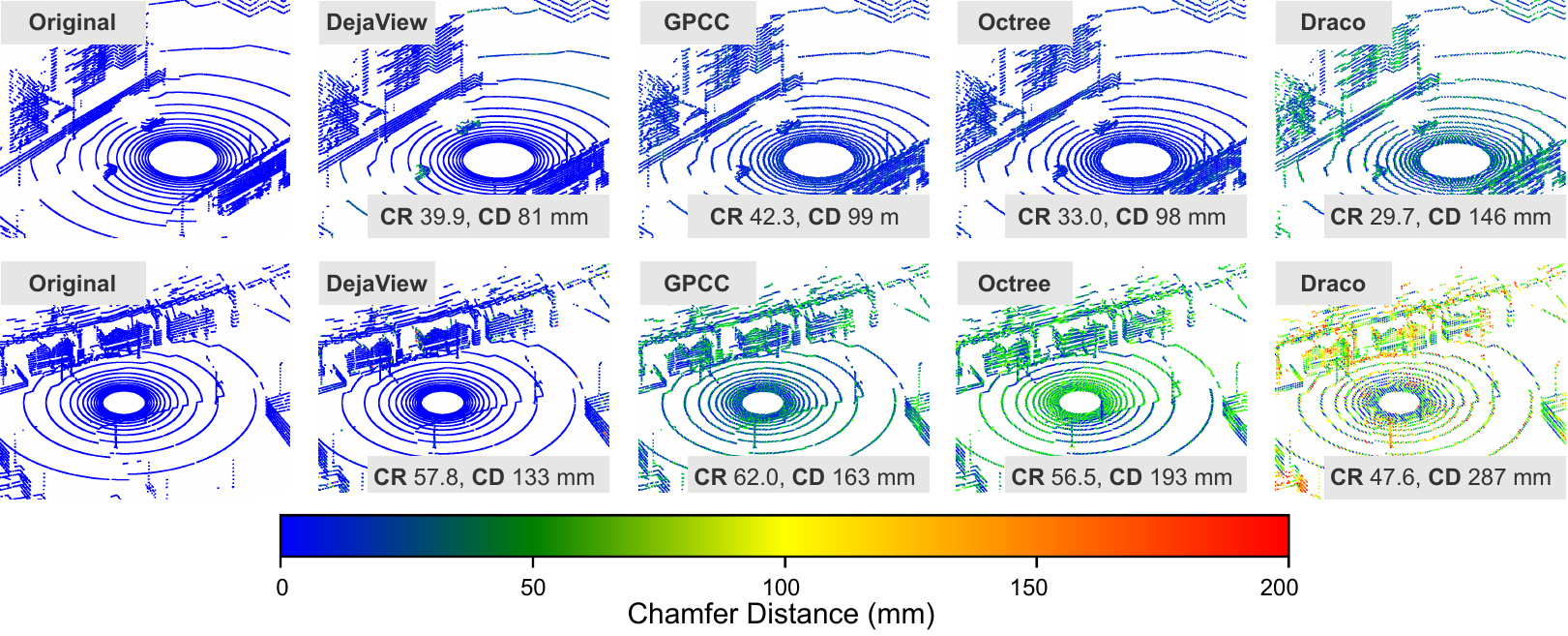}
    \caption{\sysname achieves lower reconstruction error, measured by CD, compared to GPCC, Octree, and Draco, particularly at higher CR. }
    \label{fig:qualitative_results}
    \vspace{-5pt}
\end{figure*}

\begin{figure}[t]
    \centering
    % Subfigure 1
    \begin{subfigure}{0.49\columnwidth}
    \centering
    \includegraphics[width=\columnwidth]{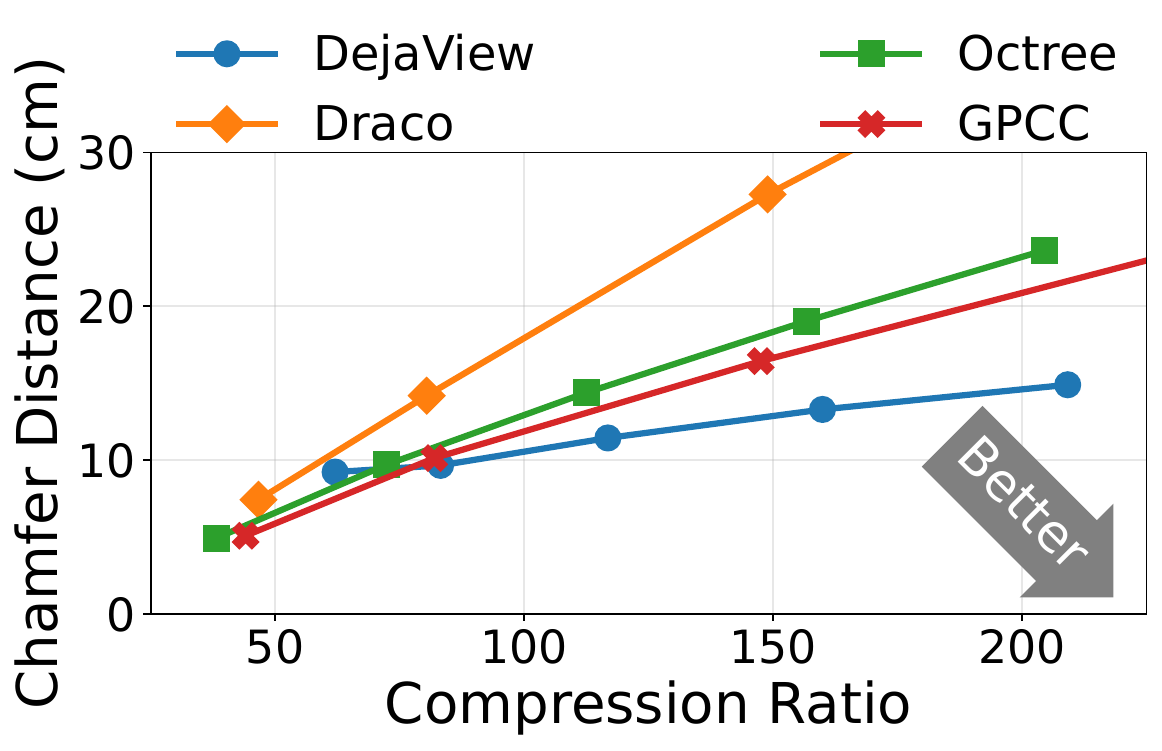} 
    \caption{\sysname ensures high compression ratios at lower Chamfer distances.}
    \label{fig:real_CD}
    \end{subfigure}
    % Subfigure 2
    \begin{subfigure}{0.49\columnwidth}
    \centering
    \includegraphics[width=\columnwidth]{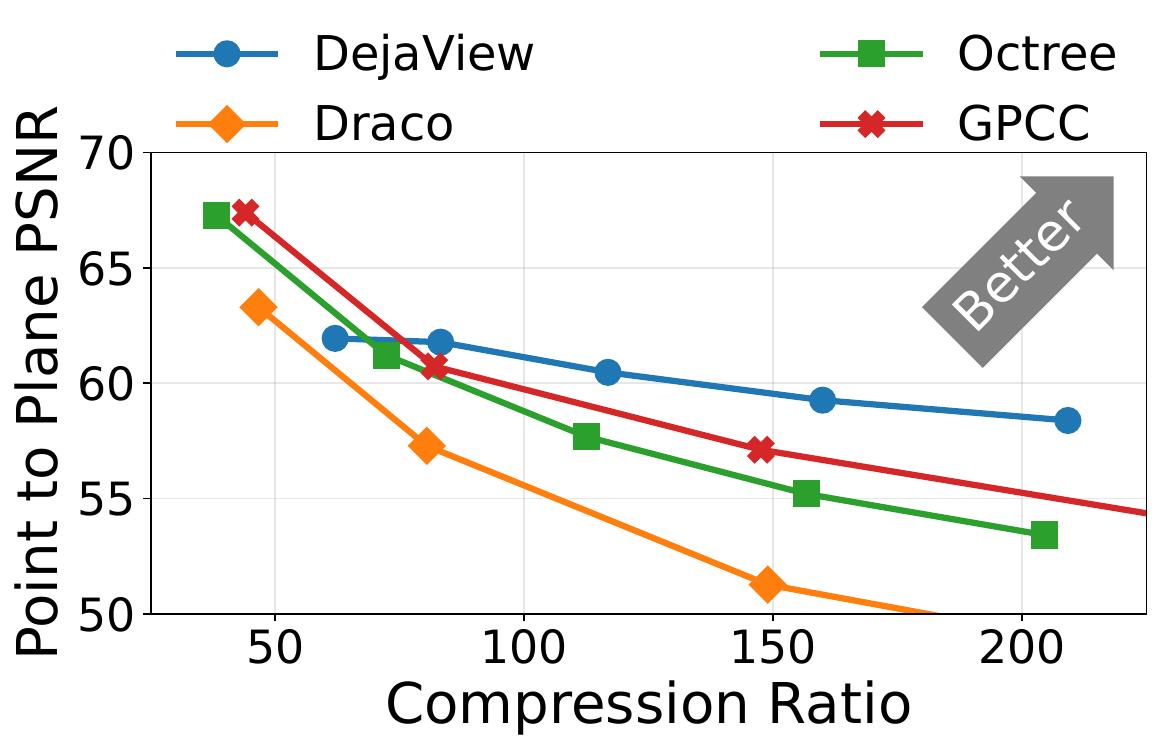} 
    \caption{\sysname achieves higher compression ratios at high PSNRs.}
    \label{fig:real_PSNR}
    \end{subfigure}
    \caption{Even at high compression ratios, \sysname is able to reconstruct the compressed point clouds accurately.}
\end{figure}

% \begin{figure}[t]
%     \centering
%     \includegraphics[width=0.7\columnwidth]{figures/end-to-end_real_CD_1.pdf} 
%     \caption{\sysname ensures high compression at low Chamfer distances.}
%     \label{fig:real_CD}
% \end{figure}

% \begin{figure}[t]
%     \centering
%     \includegraphics[width=0.7\columnwidth]{figures/end-to-end_real_PSNR_1.pdf} 
%     \caption{\sysname achieves higher compression ratios at high PSNRs.}
%     \label{fig:real_PSNR}
% \end{figure}

% \subsection{Qualitative Results }

\begin{figure*}[t]
\vspace{1em}
    % Second row
    \begin{subfigure}{0.33\textwidth}
        \centering
        \includegraphics[width=\textwidth]{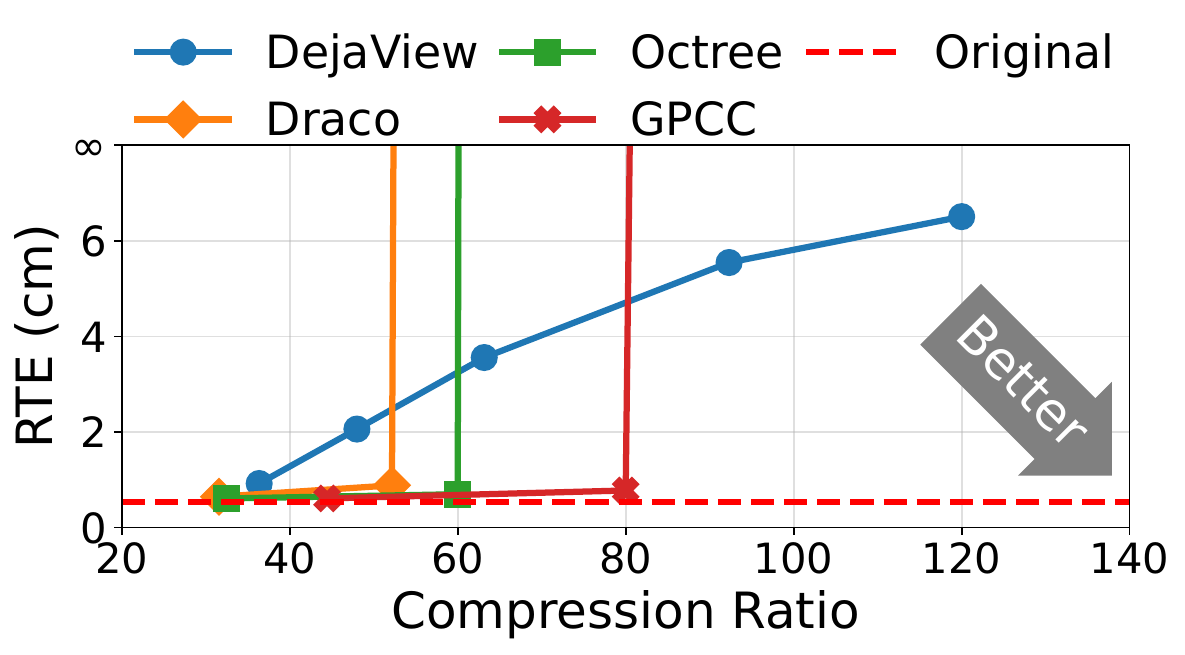}
        \caption{\sysname achieves relatively accurate localization despite high compression.}
        \label{fig:rte}
    \end{subfigure}
    \hfill
    \begin{subfigure}{0.33\textwidth}
        \centering
        \includegraphics[width=\textwidth]{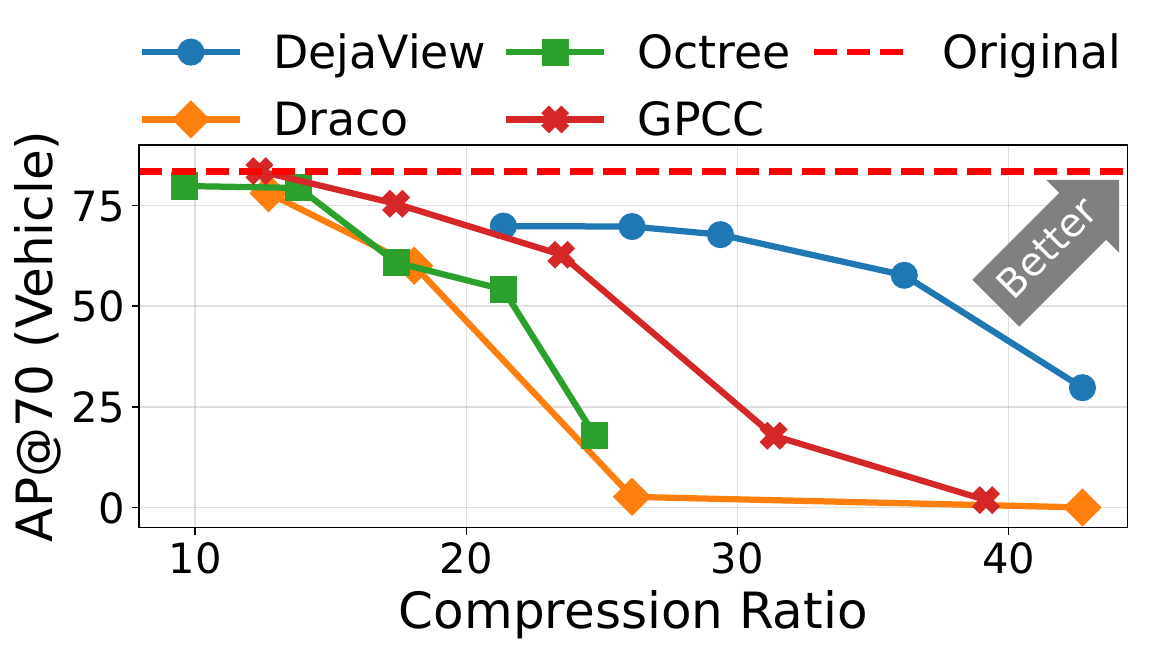}
\caption{\sysname ensures accurate object detection at high compression levels.}
        \label{fig:object-detection}
    \end{subfigure}
    \hfill
    \begin{subfigure}{0.33\textwidth}
        \centering
        \includegraphics[width=\textwidth]{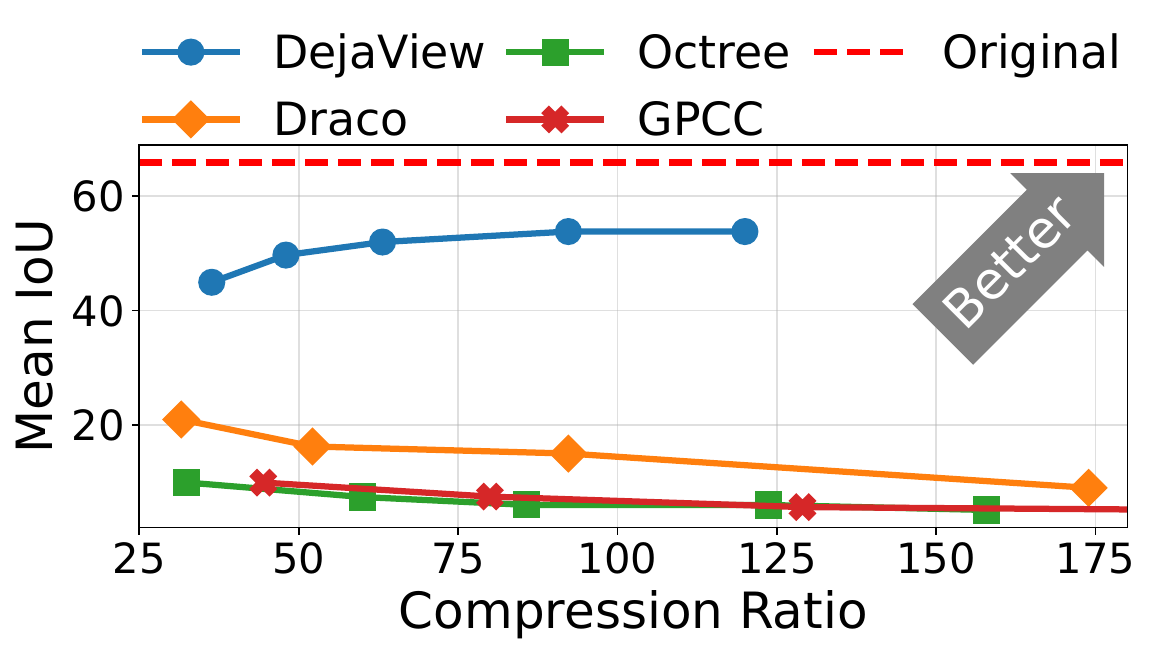}
        \caption{\sysname, despite high compression, achieves accurate segmentation.}
        \label{fig:semantic-segmentation}
    \end{subfigure}
    \caption{Performance evaluation of \sysname on perception tasks against three baselines (Draco, Octree, and GPCC).}
    % \textbf{Top row:} Compression vs. reconstruction quality on a real-world dataset. \textbf{Bottom row:} Performance comparison of \sysname, Octree, and Draco on downstream tasks in autonomous driving.}
    \label{fig:combined}
    \vspace{-10pt}

\end{figure*}

\subsection{End-to-end Experiments}

We used an end-to-end implementation of \sysname for our real-world experiments and evaluated its ability to compress point clouds
% We evaluated \sysname 
against three baselines: the Point Cloud Library's Octree-based compression~\cite{Rusu_ICRA2011_PCL}, an open-source implementation of Draco~\cite{Draco} from Google, and geometry-based point cloud compression (GPCC) from MPEG~\cite{mpeg_gpcc_software}.

We evaluated all four techniques at different levels of compression. 
All techniques have a knob that trades off compression for reconstruction error.
For \sysname, we control the amount of compression using the \textit{distance threshold}.
Draco has a parameter to set the number of bits per point. GPCC controls compression with a quantization parameter.
% in which we can set the number of bits per point. 
% A higher quantization parameter translates to higher compression. 
For Octree, we control the compression ratio using its octet resolution and point resolution.

In the first experiment, we measured the trade-off between compression ratio and reconstruction error (Chamfer distance and PSNR). 
\figref{fig:real_CD} plots compression ratio on the x-axis and the Chamfer distance on the y-axis for the four techniques using the real-world dataset. 
The ideal compression technique will have a high compression ratio for lower Chamfer distances (bottom right of \figref{fig:real_CD}). 
Draco (orange line) can compress point clouds more than Octree (green line) and GPCC (red line) but does so by trading off Chamfer distance. 
Compared to Octree, Draco, and GPCC, \sysname consistently achieves a significantly higher compression ratio at much lower Chamfer distances. 
For instance, for a 14~cm Chamfer distance, the compression ratios for GPCC, Octree, and Draco are 122, 112, and 80, respectively, whereas for \sysname, it is 220! 
\textit{Thus, \sysname can compress \sclouds 
% 2$\times$ 
80\% more than GPCC at the same Chamfer distance.}

% \ali{1.8x doesn't seem that interesting. Can we phrase it in some other way like have 54 percent more compression than Octree}
% \fawad{The percentage comes out to 180\%. So, its more or less the same.}

% \fawad{Question: Should we talk about using ICP before performing diff?}

As we increase the compression ratio for the four techniques, the Chamfer distance also increases. 
Ideally, we would like the increase in the Chamfer distance to be small. Otherwise, the reconstructed data would not be usable for downstream applications. 
By increasing GPCC's compression ratio by 140, the Chamfer distance increases by 13~cm. 
This is undesirable. 
\sysname, on the other hand, trades off only 4~cm in Chamfer distance for increasing the compression ratio by 150.
\textit{Unlike the other three techniques, \sysname can ensure high compression by trading off small amounts of Chamfer distance (reconstruction error).}

For the same experimental setup, \figref{fig:real_PSNR} plots point-to-plane PSNR as a function of the compression ratio. 
An ideal compression technique will have a high compression ratio for a high PSNR \ie it should operate on the top right of the graph. 
The results from this experiment are similar to that of the previous \ie \sysname achieves a high compression ratio for low reconstruction error (or, high PSNR). 
For a PSNR of 59, \sysname is doing approximately 
%1.5x and 2.3x
98\%, 120\% and 175\%
% \mustafa{please use small 'x' instead of large X. I fixed this during my pass so don't overwrite}
better than GPCC, Octree and Draco, respectively.
\textit{This demonstrates that \sysname can preserve both the surface geometry information of \scloud and the positional information of the points at high compression levels, while GPCC, Draco, and Octree cannot.}

We have also performed qualitative analysis by visualizing the compressed point clouds of the four techniques at different levels of compression in \figref{fig:qualitative_results}. 
We color-coded all points in the point clouds based on their distances from points in the ground-truth uncompressed point cloud on the left. 
A blue color indicating a lower Chamfer distance shows that the compression techniques preserve the 3D position of the given point with high accuracy. 
Other colors (green, yellow, and red) indicating a higher Chamfer distance show that the compression technique could not preserve the 3D position of the points. \sysname demonstrates superior capability to preserve the 3D positions of points, even at higher compression ratios, compared to GPCC, Draco, and Octree. 

While recent deep learning-based methods like OctAttention~\cite{fu2022octattention} can achieve high compression ratios,  we exclude them from end-to-end experiments due to their significantly higher compression and decompression times. Instead, we analyze them separately in \secref{ss::learning_based_methods}.

% \begin{figure*}[t]
%     \centering
%     % Subfigure 1
%     \begin{subfigure}{0.33\textwidth}
%         \centering
%         \includegraphics[width=\textwidth]{figures/RTE_1.pdf}
%         \caption{Localization}
%         \label{fig:rte}
%     \end{subfigure}
%     % Subfigure 2
%     \begin{subfigure}{0.33\textwidth}
%         \centering
%         \includegraphics[width=\textwidth]{figures/mAP_1.pdf}
%         \caption{3D Object Detection}
%         \label{fig:object-detection}
%     \end{subfigure}
%     % Subfigure 3
%     \begin{subfigure}{0.33\textwidth}
%         \centering
%         \includegraphics[width=\textwidth]{figures/IOU_1.pdf}
%         \caption{3D Semantic Segmentation}
%         \label{fig:semantic-segmentation}
%     \end{subfigure}
%     \caption{Performance comparison of \sysname, Octree, and Draco on downstream tasks in autonomous driving. }
%     \label{fig:sensitivity_analysis}
% \end{figure*}

\begin{figure*}[t]
    \centering
    % Subfigure 1
    \begin{subfigure}{0.31\textwidth}
        \centering
        \includegraphics[width=\textwidth]{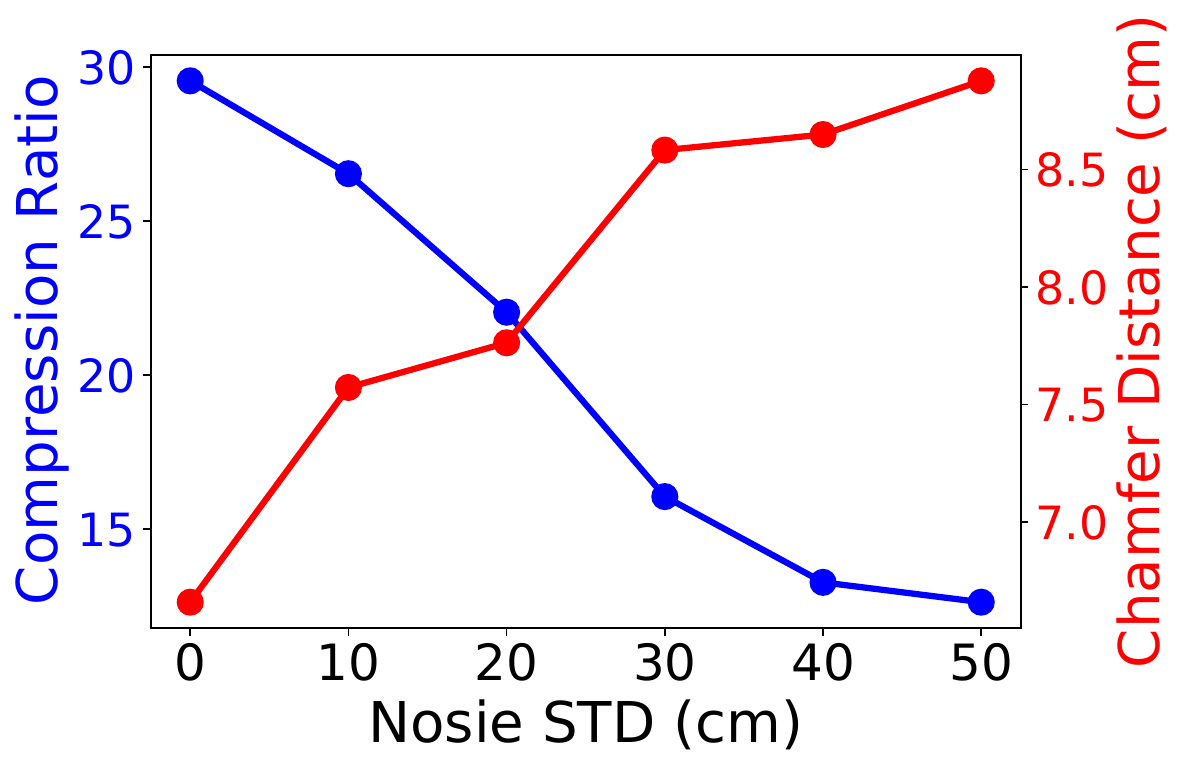}
        \caption{Sensor Noise}
        \label{fig:noise}
    \end{subfigure}
    % Subfigure 2
    \hfill
    \begin{subfigure}{0.31\textwidth}
        \centering
        \includegraphics[width=\textwidth]{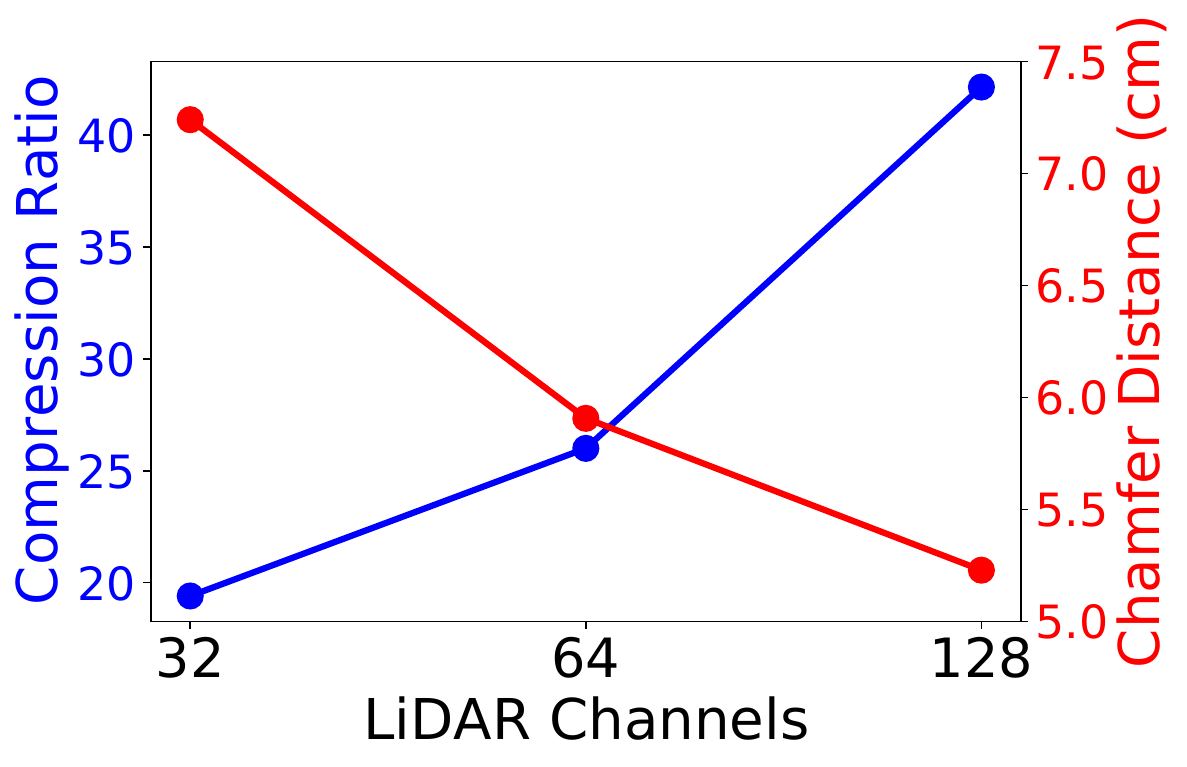}
        \caption{LiDAR Channels}
        \label{fig:channels}
    \end{subfigure}
    % Subfigure 3
    \hfill
    \begin{subfigure}{0.31\textwidth}
        \centering
        \includegraphics[width=\textwidth]{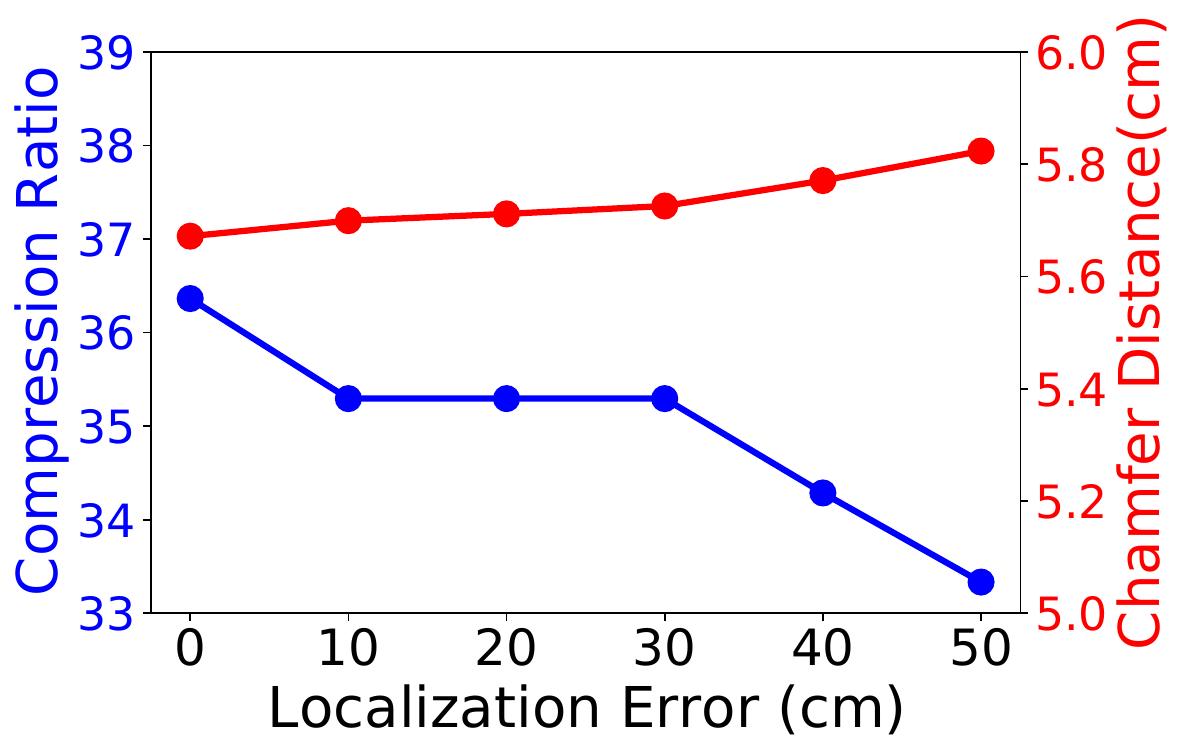}
        \caption{Localization Error}
        \label{fig:localization_error}
    \end{subfigure}
    % % Subfigure 4
    % \begin{subfigure}{0.24\textwidth}
    %     \centering
    %     \includegraphics[width=\textwidth]{figures/distance-threshold.pdf}
    %     \caption{Distance Threshold}
    %     \label{fig:distance_threshold}
    % \end{subfigure}
    \caption{Sensitivity Analysis of \sysname}
    \label{fig:sensitivity_analysis}
    \vspace{-10pt}
\end{figure*}

\subsection{Application-level Results}
While Chamfer distance and PSNR are similarity metrics, they do not quantify the impact on downstream applications. AV-captured point clouds are crucial for training and testing perception modules such as localization, object detection, and semantic segmentation. 
This section evaluates \sysname's compression effects on these critical AV tasks.

\parab{Localization. }
Localization determines the precise pose of an AV within a 3D map.
AVs use NDT~\cite{biber2003normal} to align the point cloud of the current frame with the 3D map to estimate their pose. 
We evaluated the precision of localization using the relative translation error (RTE)~\cite{rte}, the root mean square distance between the ground truth and the estimated positions.

For this experiment, we used the synthetic dataset generated from CARLA, with which we have ground-truth 3D positions. 
Using data from a single day, we build a 3D map and then localize five point cloud sets on this map: raw uncompressed and those compressed by GPCC, Octree, Draco, and \sysname.

% \begin{figure}[t]
%     \centering
%     \includegraphics[width=\columnwidth]{figures/RTE.pdf} 
%     \caption{Even at high levels of compression, the localization error (RTE) for \sysname's point clouds is within 7~cm.}
%     \label{fig:rte}
% \end{figure}

% \figref{fig:rte} shows the localization error (RTE) against the compression ratio for \sysname (blue), Draco (orange), and Octree (green). 
% The dotted red line shows the RTE for raw point clouds. 
% An ideal compression scheme will have a high compression ratio and low localization (RTE), \ie bottom right of \figref{fig:rte}.
% As the compression ratio increases, so do the Chamfer distance and localization errors.
% Higher Chamfer distances indicate a large deviation of compressed point positions from the raw data.
% Although NDT can accurately align raw point clouds, the deviation in point positions for compressed point clouds leads to higher localization errors.
% Of the three schemes, only \sysname's localization error remains below 7~cm even with compression ratios up to 120. 
% NDT cannot localize point clouds at all after a compression ratio of 50 and 60 for Draco and Octree, respectively.
% \textit{These results show that \sysname can achieve higher compression with minimal effect on end-to-end localization accuracy.}

\figref{fig:rte} shows the localization error (RTE) against the compression ratio for \sysname (blue), GPCC (orange), Octree (red), and Draco (green).
The dotted red line shows the RTE for raw point clouds. 
An ideal compression scheme will have a high compression ratio and low localization error (RTE), \ie bottom right of \figref{fig:rte}.
As the compression ratio increases, so do the Chamfer distance and localization errors.
Higher Chamfer distances indicate a large deviation of compressed point positions from the raw data.
Although NDT can accurately align raw point clouds, the deviation in point positions for compressed point clouds leads to higher localization errors.
Of the four schemes, only \sysname's localization error remains below 7~cm even with compression ratios up to 120. 
NDT cannot localize point clouds at all after a compression ratio of 50, 60, and 80 for Draco, Octree, and GPCC, respectively.
\textit{These results show that \sysname can achieve higher compression with minimal effect on end-to-end localization accuracy.}

\parab{Object Detection. }
Object detection is a crucial component in the autonomous driving pipeline. 
We evaluated the effect of compression on the performance of a 3D object detection model, Part-\textit{A$^{2}$} net~\cite{shi2020points}. 
We trained this model for 100 epochs using 2,700 point clouds from CARLA. 
For testing, we used 3 days of data from CARLA, 
from which the first day's data were used to build a 3D map and the second day's data for a reference dataset. 
We compressed the third day data using \sysname, GPCC, Octree, and Draco.

% \begin{figure}[t]
%     \centering
%     \includegraphics[width=\columnwidth]{figures/mAP.pdf} 
%     \caption{Average Precision (AP) of reconstructed point clouds at IoU threshold of 70 for vehicles on 3D object detection task.}
%     \label{fig:object-detection}
% \end{figure}

\figref{fig:object-detection} plots average precision (AP) at an intersection-over-union (IoU) 
threshold of 70 (AP@70) for vehicle bounding boxes as a function of compression ratio. 
AP@70 is the area under the precision-recall curve. A higher value of AP indicates better overall detection performance~\cite{everingham2010pascal}.
Ideally, a compression scheme should have high AP for high compression ratio \ie top-right of \figref{fig:object-detection}. 
As expected, with higher compression ratios, the AP drops. 
However, this drop is rapid for Octree, GPCC, and Draco. 
On the other hand, \sysname's AP degrades gracefully with an increasing compression ratio. 
At AP values of 30, \sysname compresss
% 5x and 1.5x 
95\%, 79\%, and 48\%
more than Draco, Octree, and GPCC, respectively.

\parab{3D Semantic Segmentation. }
3D semantic segmentation is another important task in the autonomous driving pipeline. In this task each point in the point cloud is assigned a semantic label like road, car, tree, \etc.
% To evaluate compression effects on semantic segmentation 
(assigning object labels to individual points), 
we fine-tuned the last 9 layers of a MinkowskiNet-based model~\cite{choy20194d, openpcseg2023} pre-trained on the Semantic KITTI dataset~\cite{behley2019semantickitti}.
For this, we evaluated mean IoU (mIoU), across all class labels, a metric that determines the precision of semantic segmentation (formally defined in~\cite{behley2019semantickitti}). 
A technique that assigns correct labels to every point in the point cloud will have a mIoU of 100. 
Thus, a higher mIoU is better.

% \begin{figure}[t]
%     \centering
%     \includegraphics[width=\columnwidth]{figures/IOU.pdf} 
%     \caption{Mean Intersection over Union (IoU) performance of reconstructed point clouds on 3D semantic segmentation task.}
%     \label{fig:semantic-segmentation}
% \end{figure}

\figref{fig:semantic-segmentation} shows IoU performance at various compression ratios.
An ideal compression technique will have high mIoU for high compression ratio \ie it will operate to the top right of \figref{fig:semantic-segmentation}. 
Of these four schemes, \sysname's mIoU scores are significantly better for all compression ratios. 
Octree, GPCC, and Draco have low mIoU even at lower compression ratios and it further decreases as the compression ratio increases.
% \textcolor{blue}{
At a compression ratio of 122, \sysname achieves a 5X higher mean IoU than Draco, while Octree and GPCC exhibit nearly zero mean IoU at this compression ratio.
% }
Interestingly, \sysname's performance improves at higher compression ratios because of the reduced impact of Draco compression.

\begin{figure*}[t]
    \centering
    % First row
    \begin{subfigure}{0.64\textwidth}
        \centering
        \includegraphics[width=\textwidth]{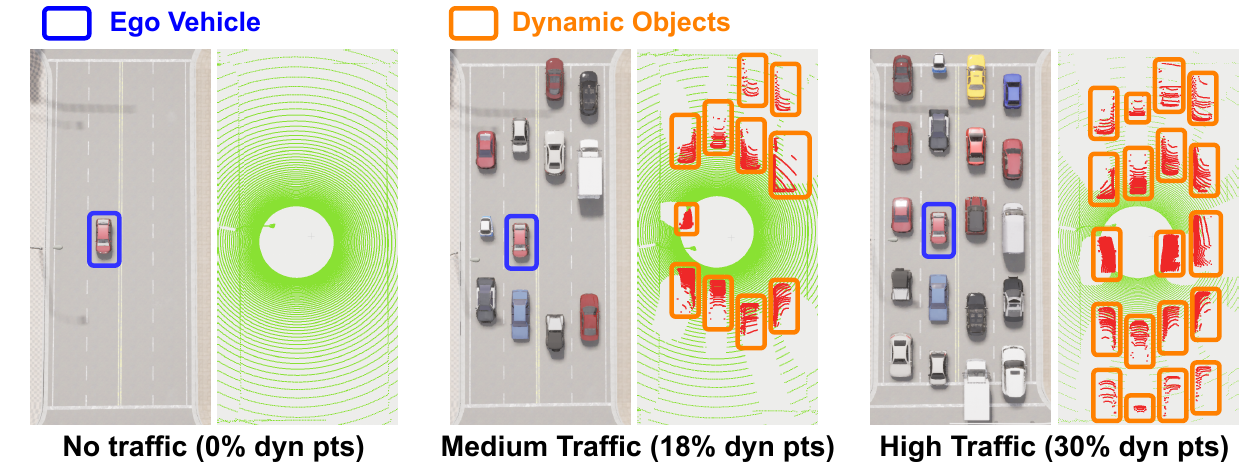}
           \caption{Bird-eye-view CARLA images and point clouds under different traffic conditions. The dynamic points peak at 30\% when the ego-vehicle is fully surrounded by nearby vehicles.}
           \label{fig:traffic_condition}
    \end{subfigure}
    % \hspace{10pt}
    \hfill
    \begin{subfigure}{0.35\textwidth}
        \centering
        \includegraphics[width=\textwidth]{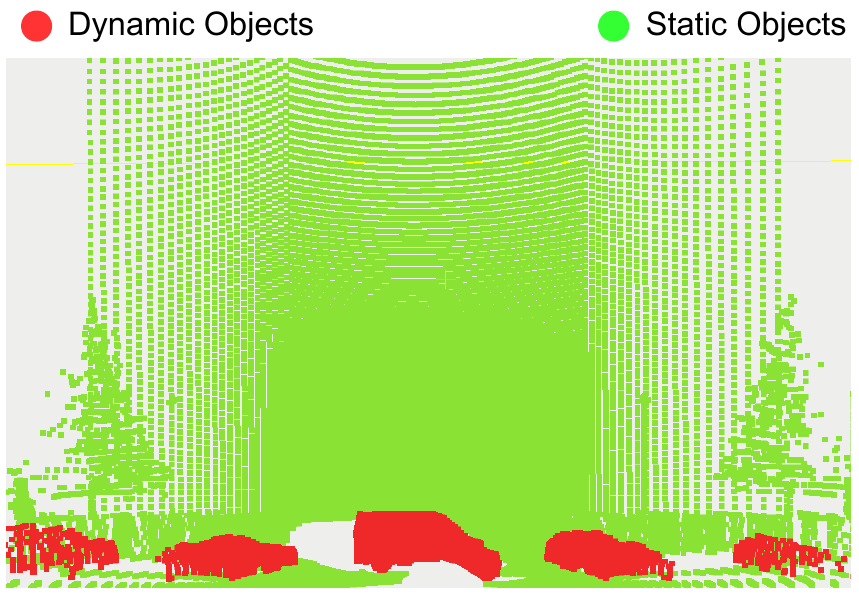}
        \caption{2D rendering of a LiDAR's vertical field-of-view shows dynamic objects only occupy a small portion of the point cloud.}
        \label{fig:vertical-fov}
    \end{subfigure}
    
     \caption{Illustrations of traffic scenarios generated in CARLA for evaluating the effect of dynamic content percentage on compression ratio of \sysname.}
    \label{fig:dynamic_environment}
    \vspace{-5pt}

\end{figure*}

\subsection{Sensitivity Analysis}
\label{ss:sensitivity_analysis}

\parab{Sensor Noise. }
Real-world LiDAR data inherently contain sensor noise, which shifts 3D points from their actual positions. 
The magnitude of this noise varies, typically ranging from 1~cm to 10~cm for commercial-off-the-shelf LiDAR sensors~\cite{lidarnoise_holzhuter2023technical}. 
To evaluate the impact of sensor noise on \sysname, we simulated different noise levels in the point cloud using CARLA's Gaussian noise model.
We varied the standard deviation of LiDAR noise from 0 to 50~cm. 
In \figref{fig:noise}, we plot the compression ratio (blue) and the Chamfer distance (red) as functions of the sensor noise for \sysname with the \textit{distance threshold} set at 10~cm. 

As the standard deviation of sensor noise increases from 0~cm to 50~cm , the compression ratio decreases from 30 to 13.
At higher noise levels, the 3D points in \scloud deviate from their actual positions. 
Consequently, \sysname cannot associate them with points in \tcloud. 
As a result, these are instead stored as exclusive points. 
As noted before, increased exclusive points lead to a lower compression ratio. 
% Moreover, the Chamfer distance also increases with increased noise. 
Moreover, this also increases the Chamfer distance from 6.6~cm to 8.9~cm. This occurs because more points are classified as exclusive and applying Draco compression to these points reduces their precision.
In turn, this increases the Chamfer distance.
% This is because a higher proportion of points are classified as exclusive; when Draco compression is applied to these points, their precision decreases, leading to larger chamfer distances. 

To quantify Draco's impact on the Chamfer distance, we compressed 2000 point clouds using \sysname with a 10~cm \textit{distance threshold}.
When decompressing the point clouds, we separated the points into those compressed by Draco and those stored as indices. 
Draco compressed points had a $3.3$x higher Chamfer distance compared to those stored as indices. 
Hence, a higher number of exclusive points reduces both the compression ratio and the overall quality of the reconstruction.

% \begin{figure}[t]
%     \centering
%     \includegraphics[width=\columnwidth]{figures/noise.pdf} 
%     \caption{\sysname performs best at low-levels of sensor noise \ie high compression and low Chamfer distance.}
%     \label{fig:sensor-noise}
% \end{figure}

% \begin{figure}[t]
%     \centering
%     \includegraphics[width=\columnwidth]{figures/channels.pdf} 
%     \caption{Effect of LiDAR channels on compression ratio and chamfer distance for \sysname}
%     \label{fig:channels}
% \end{figure}

\parab{LiDAR Channels. }
The number of channels (laser beams) in a LiDAR sensor impacts the point density (points per unit area) of the collected point clouds.
To evaluate the effect of the number of LiDAR beams on \sysname, we generated CARLA data for 32, 64, and 128 beam LiDARs. 
In \figref{fig:channels}, we plot the compression ratio (blue) and Chamfer distance (red) of \sysname as functions of the LiDAR channels.
As the number of channels increases from 32 to 64, the compression ratio of \sysname also increases from 19 to 42.
This is because, with denser point clouds \sysname has a higher probability of finding common points (within the \textit{distance threshold}) between \scloud and \tcloud. 
This translates to fewer exclusive points and a higher compression ratio. 
Consequently, fewer exclusive points are subject to Draco's quantization, reducing Chamfer distance by 2~cm.

\parab{Localization Error. }
Localization is a process in which AV uses sensor data to find its precise location in the world. This location information is then associated with the corresponding sensor data. As explained in \secref{s:design}, \sysname uses this location information to find a \tcloud for each \scloud. To evaluate the effect of localization error on \sysname performance, we generated data from CARLA along with the ground-truth location of each point cloud. To simulate the localization error, we added a random number sampled from the uniform distribution to the ground-truth location. The range of the uniform distribution was changed to control the magnitude of the error. \figref{fig:localization_error} plots the compression ratio (blue) and the Chamfer distance (red) of \sysname versus the localization error. A localization error of 50~cm only reduces the compression ratio from 36 to 33 and increases the Chamfer distance by 0.1~cm. The slight reduction in compression ratio is because localization error can lead to suboptimal selection of a \tcloud for each \scloud. This, in turn, means fewer common points, thus affecting the compression ratio negatively.

% \ali{Shepherd + reviewers: 
% Discussion on robustness in dynamic environments
% No discussion on robustness in dynamic environments. The paper only categorizes dynamic environments based on “traffic volume,” without considering the motion characteristics of dynamic objects (e.g., fast-moving vehicles, sudden crossing pedestrians) and their impact on compression performance. When dynamic objects cause the proportion of exclusive points to exceed 50\%, can DejaView’s compression ratio still outperform traditional methods? Such extreme dynamic scenarios were not included in the experiments, limiting the generalizability of the results. (Rev E)
% CarLA is a sophisticated simulator based on real-world physics, but it cannot fully represent actual driver behavior or human movement. Relying on simulation for dynamic environments is somewhat unrealistic. Adding experiments conducted with real-world data would make the evidence more compelling. (RevD)
% }

\parab{Dynamic Environment. } 
% An AV's enviroment consists of both static elements (roads, buildings) and dynamic elements (vehicles, pedestrians).
% and dynamic entities, such as cars and pedestrians. 
In this experiment, we systematically analyze the effect of the dynamic objects (vehicles, pedestrians, and cyclists \etc) on \sysname's compression ratio. 
Because it is difficult to conduct controlled experiments in the real-world with dynamic objects, for safety reasons and otherwise, we use CARLA for these evaluations.
% To systematically analyze the effect of dynamic objects on compression ratio, we use CARLA to simulate different amounts of traffic density. 
% This is difficult to do in real-world experiments which are difficult to reproduce in real-world experiments. 
We generate three CARLA scenarios (\figref{fig:traffic_condition}) with varying number of dynamic objects. 
The \textit{no-traffic} scenario contains 0\% dynamic points, \textit{medium traffic} contains 18\% dynamic points, and \textit{high traffic} contains 30\% dynamic points.
% As shown in~\figref{fig:traffic_condition}, we generate three CARLA scenarios with different numbers of dynamic objects: a high-traffic scenario with 15 to 20 vehicles that contribute about 30\% dynamic points per frame, a medium-traffic scenario with 10 to 15 vehicles that contribute around 18\% dynamic points, and a no-traffic scenario where all points belong to static objects. 
Even in \textit{high-traffic} scenarios, where the ego-vehicle is completely surrounded by dynamic objects, dynamic points only occupy 30\% of the point cloud. This is because, although the vehicle is surrounded by objects along the horizontal field of view of LiDAR, these objects only occupy a small area on the vertical field of view~(\figref{fig:vertical-fov}).
% limited portion of the LiDAR’s vertical field of view.

\begin{figure}[t]
    \vspace{-5pt}

    \centering
    \includegraphics[width=0.6\columnwidth]{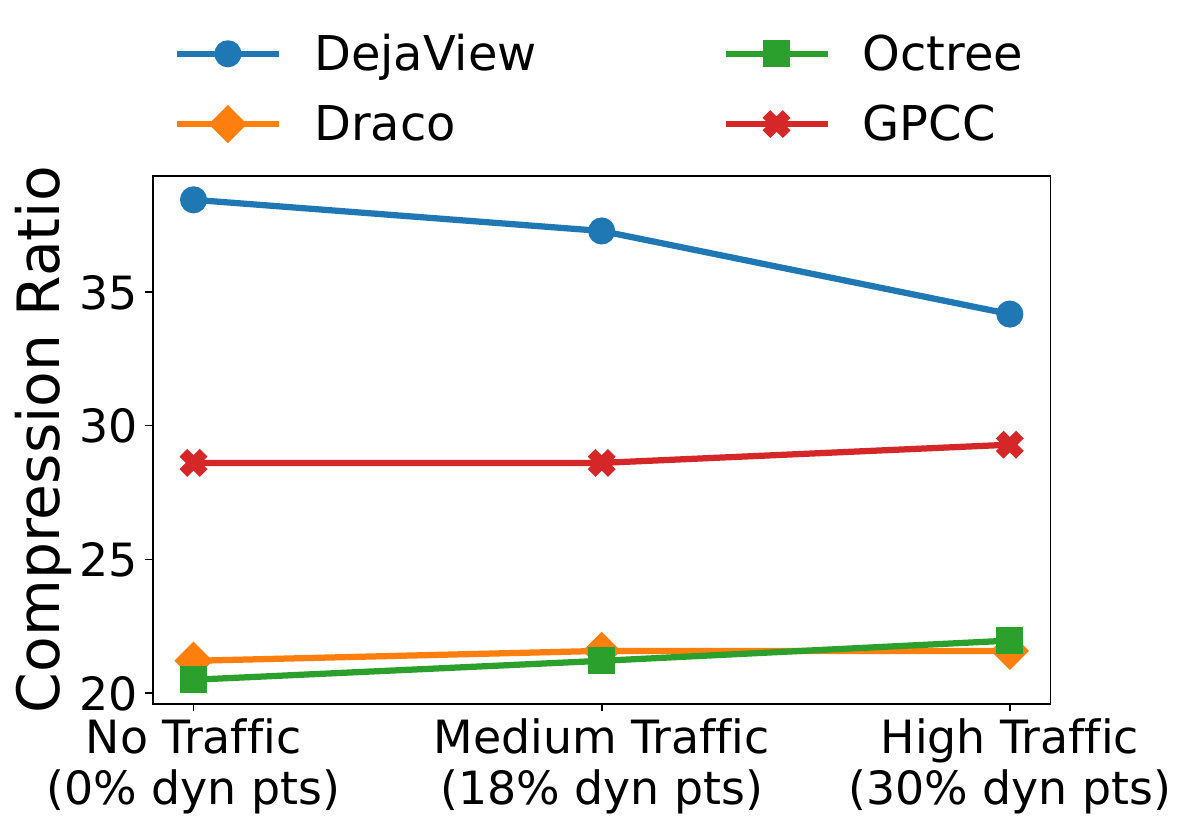} 
    \caption{\sysname ensures high compression across all traffic conditions.}
    % with different percentages of dynamic points.}}    
    \label{fig:dynamic_content}
\end{figure}

\figref{fig:dynamic_content} shows the effect of the number of dynamic objects on the compression ratio of \sysname and other baselines.
\sysname consistently outperforms the three baselines.
At lower traffic densities, \sysname achieves higher compression ratios. 
This happens because fewer dynamic objects result in fewer exclusive points in \scloud relative to \tcloud, which increases compression ratio for \sysname.
% \sysname’s compression ratio. 
In contrast, GPCC, Octree, and Draco show nearly constant compression ratios since they depend on the total number of points rather than the underlying scene dynamics.

% The environment observed by the autonomous vehicle (AV) includes both static elements, such as buildings and roads, and dynamic entities, such as cars and pedestrians. To assess the impact of dynamic content on compression ratios, we generated data using CARLA in three distinct traffic scenarios: \textit{high traffic}, where the AV is surrounded by 15 to 20 cars; \textit{medium traffic}, with 10 to 15 surrounding cars; and \textit{low traffic}, featuring 6 to 10 surrounding cars during the AV's journey. \figref{fig:dynamic_content} illustrates the impact of environmental dynamics on \sysname, GPCC, Octree, and Draco for these traffic scenarios. 

% Across all scenarios, \sysname outperforms the three baselines, with the performance gap widening as the dynamic content in the environment decreases. This compression gain is because lower dynamic content results in fewer exclusive points in \scloud relative to \tcloud, thus increasing \sysname's compression ratio. In contrast, the compression ratios of GPCC, Octree and Draco remain constant, as they are solely dependent on the total number of points in the point clouds, irrespective of the environment's underlying dynamics.

% \ali{added a small section for distance threshold}

\parab{Distance Threshold. }
For two points in the source and reference clouds to be considered identical, they must lie within a certain 3D distance of one another (\textit{distance threshold}). 
\sysname uses \textit{distance threshold} as a knob to control the tradeoff between compression ratio and reconstruction error. 
By increasing the \textit{distance threshold} from 10~cm to 50~cm, \sysname trades off only 7~cm in Chamfer distance (or 5 units of PSNR) to improve the compression ratio from 30 to 66 (\figref{fig:distance_threshold}). 
Using this knob, system designers can tune \sysname for diverse application requirements \ie higher compression for applications that can tolerate low reconstruction quality, and vice versa.

% trading off only x~cm in Chamfer distance, 
% \figref{fig:distance_threshold} plots compression ratio and reconstruction error (PSNR) as a function of \textit{distance threshold}. 
% Increasing the threshold from 10~cm to 50~cm, \sysname increases compression ratio from 30 to 70 by tradeing off PSNR from 53 to 48.
% Increasing the distance threshold classifies more points as identical, improving the compression ratio at the cost of a slight reduction in reconstruction quality (PSNR).
% slightly reduces reconstruction quality (lower PSNR).
% Increasing the distance 

% does not talk about qualitative and quantitative results of compressed size. 
% does not talk about the effect ofthe  distance threshold on point count quantitatively. 
% does not talk about the effect of distnace threshold on compressed size at all. 

% \ali{modified the section below because of the following reasons:

% does not talk about the qualitative and quantitative results of compressed size. 

% does not talk about the effect of the distance threshold on point count quantitatively. 

% does not talk about the effect of distnace threshold on compressed size at all. 

% The original text can be found in the comments below.
% }

\begin{figure}[t]
    \centering
     % Subfigure 1
    \begin{subfigure}{0.49\columnwidth}
        \centering
        \includegraphics[width=\textwidth]{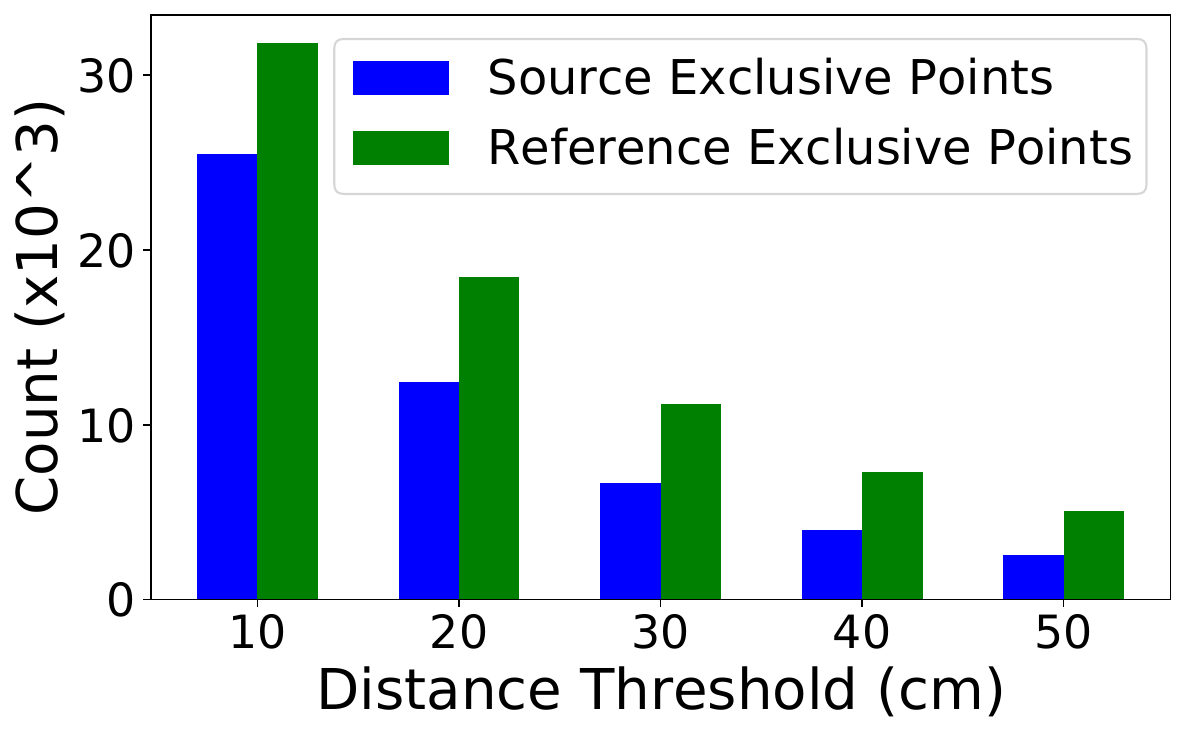}
        \caption{Point Count Breakdown}
        \label{fig:points_breakdown}
    \end{subfigure}
    % Subfigure 2
    \begin{subfigure}{0.49\columnwidth}
        \centering
        \includegraphics[width=\textwidth]{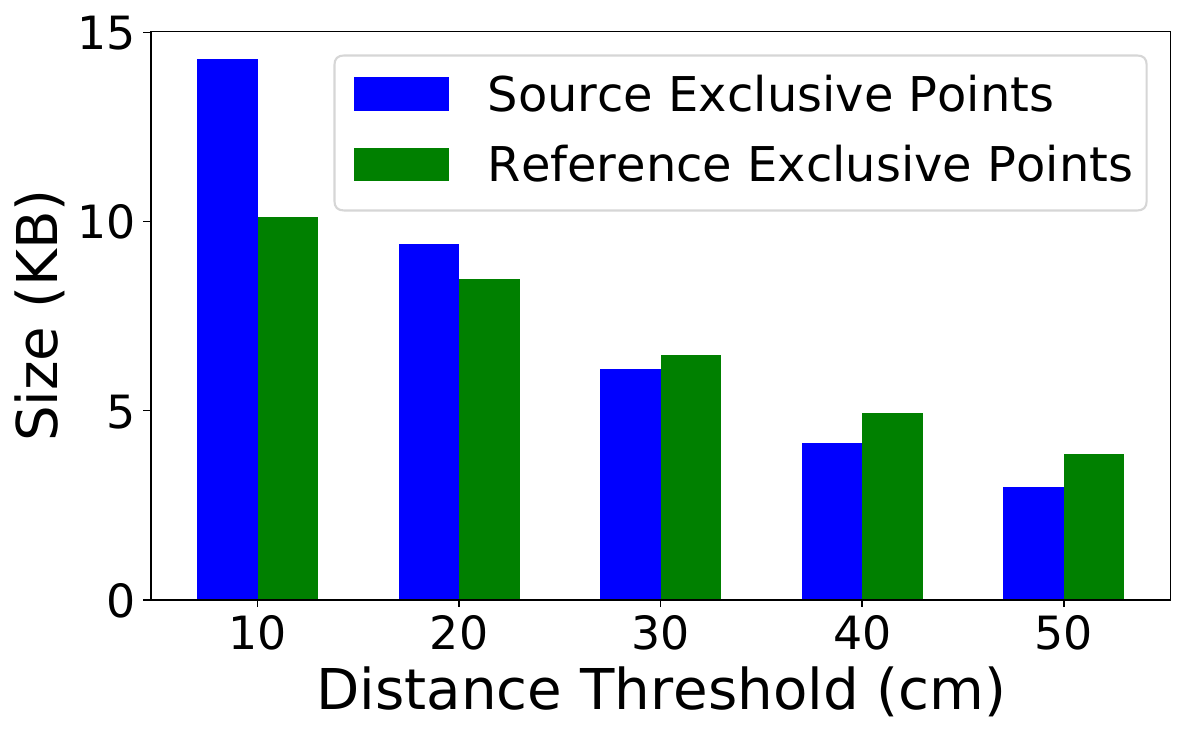}
        \caption{Size Breakdown}
        \label{fig:size_breakdown}
    \end{subfigure}

    \caption{Comparison of point count and size breakdown of exclusive points from the source and reference clouds in \sysname's compressed representation. }
    \label{fig:compressed-representation}

\end{figure}

\parab{\sysname Compressed Representation. }The compressed representation produced by \sysname consists of exclusive points from \scloud and pointers to exclusive points in \tcloud. 
\figref{fig:points_breakdown} and \figref{fig:size_breakdown} provide a breakdown for the exclusive point count and corresponding compressed size as a function of the distance threshold. 
Both \scloud and \tcloud contain 76K points. At a distance threshold of 10~cm, \sysname identifies 25K exclusive points in the \scloud and 30K exclusive points in the \tcloud.
% and 30K exclusive points in \scloud and \tcloud, respectively.
The compressed representation of these points collectively requires only 23~KB of memory.
Increasing the distance threshold to 50~cm further reduces the number of exclusive points, lowering the collective compressed size to only 7~KB. 
% While a higher distance threshold improves the compression ratio by reducing the number of exclusive points, it also results in lower reconstruction quality.

To reconstruct the \scloud, \sysname stores pointers to exclusive points in the \tclouds.
% which imposes a limit on its compression efficiency.
These pointers are the indices of exclusive points in the \tclouds. 
A single pointer is stored as an int32. 
A raw 3D point, on the other hand, would require 12 bytes instead.
% requiring only 4 bytes per index (int32) compared to 12 bytes for storing full 3D coordinates.
Finally, \sysname applies LZMA compression to the sequence of indices to further compress them.
% While LZMA compression is applied to these index sequences, there is potential to further enhance \sysname’s compression ratio by adopting more advanced techniques for storing pointers to exclusive points in \tclouds}

% The compressed representation produced by \sysname consists of exclusive points from \scloud and pointers to exclusive points in \tclouds ( \figref{fig:size_breakdown} and \figref{fig:points_breakdown}).
% % illustrate the contribution of these exclusive points in terms of size and count, respectively.
% By leveraging a cascaded diff operation, \sysname significantly reduces the number of points in \scloud from 76K to just 25K, even with a low distance threshold of 10 cm (\figref{fig:points_breakdown}). As the threshold increases, the number of exclusive points in \scloud further decreases, improving the compression ratio. However, to enable the reconstruction of \scloud from the compressed representation, \sysname must store pointers to exclusive points in \tclouds, which imposes a limit on its compression efficiency.

% Currently, \sysname stores these pointers as indices of exclusive points in \tclouds, requiring only 4 bytes per index (int32) compared to 12 bytes for storing full 3D coordinates. While LZMA compression is applied to these index sequences, there is potential to further enhance \sysname’s compression ratio by adopting more advanced techniques for storing pointers to exclusive points in \tclouds.

% \begin{figure}[t]
%     \centering
%     \includegraphics[width=\columnwidth]{figures/dynamic_env.pdf} 
%     \caption{Effect of change in dynamic content on compression ratio and chamfer distance for \sysname}
%     \label{fig:channels}
% \end{figure}

\subsection{Latency and Throughput Evaluations}
\begin{figure}[t]
% \vspace{-18pt}
    \centering
    % Subfigure 1
    \begin{subfigure}{0.49\columnwidth}
        \centering
        \includegraphics[width=\textwidth]{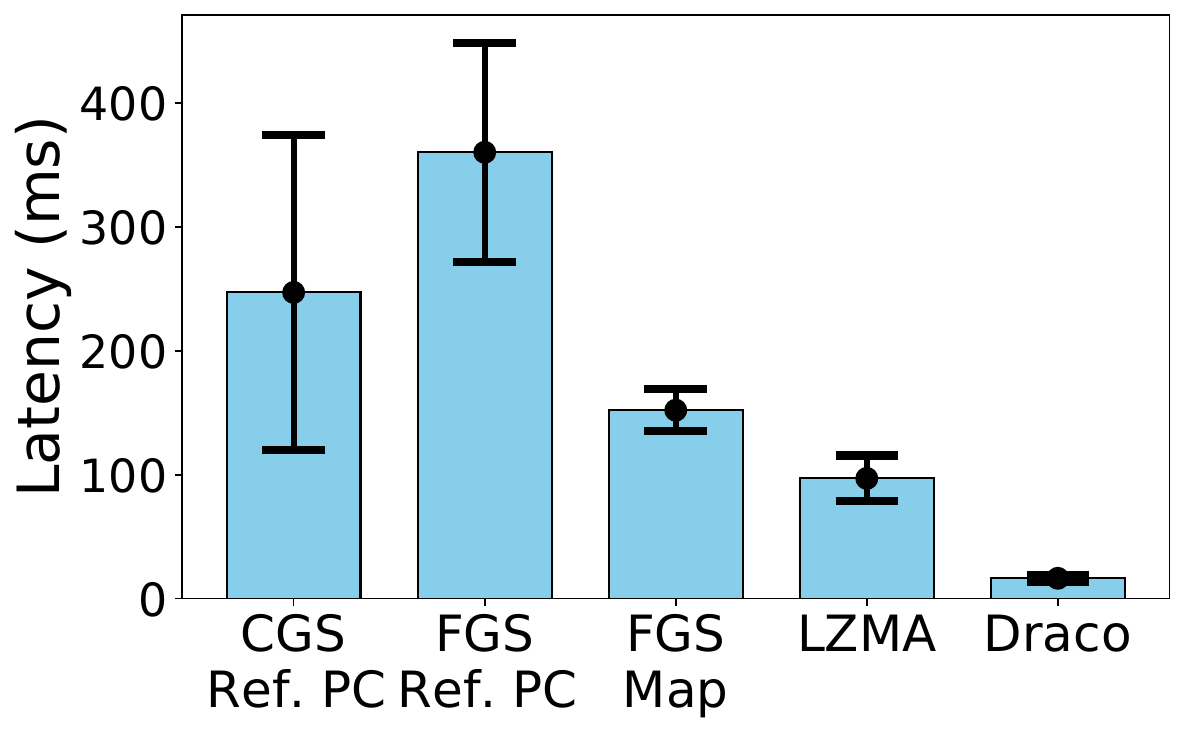}
        \caption{Latency Breakdown}
        \label{fig:latency_breakdown}
    \end{subfigure}
    % Subfigure 2
    \begin{subfigure}{0.49\columnwidth}
        \centering
        \includegraphics[width=\textwidth]{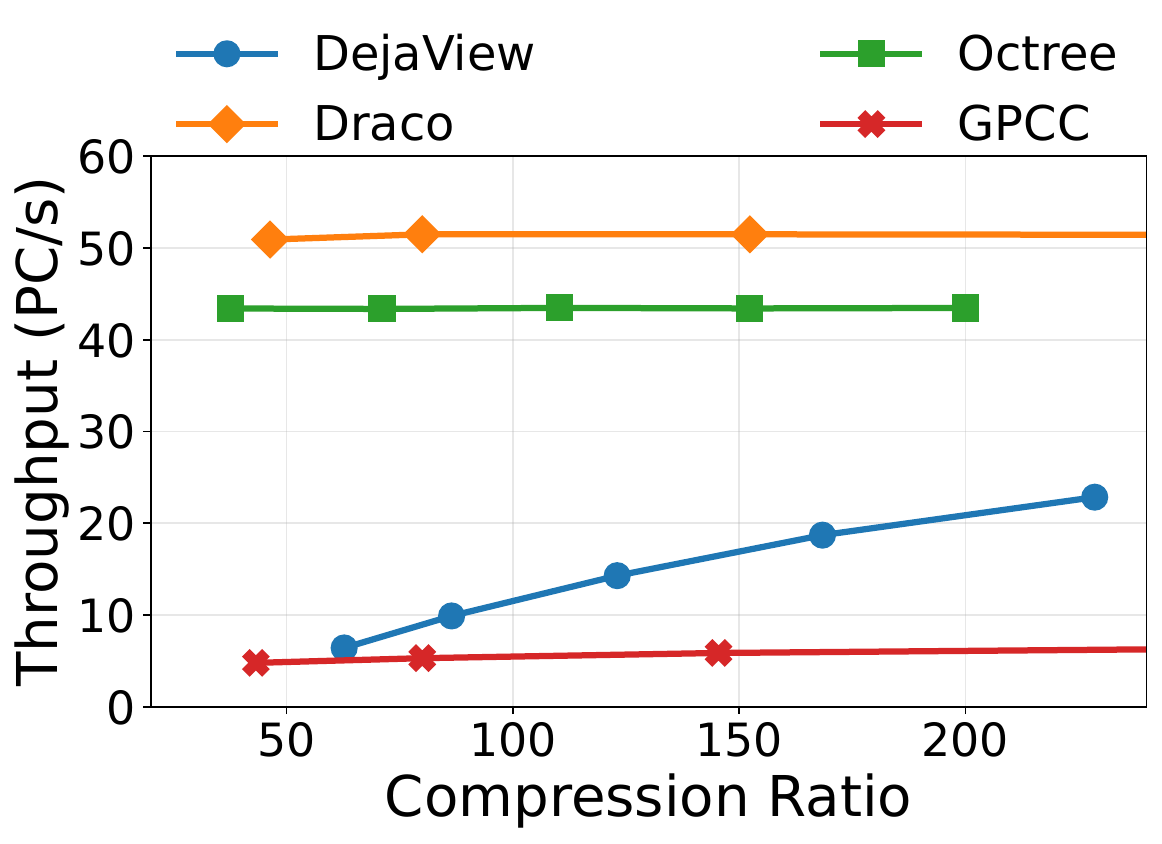}
        \caption{Throughput}
        \label{fig:throughput}
    \end{subfigure}
   
    \caption{(a) Latency breakdown of different components of \sysname.
(b) End-to-end throughput of \sysname increases with compression ratio, while it remains almost constant for all other compressors.}
\end{figure}

\figref{fig:latency_breakdown} shows the latency breakdown per point cloud across different components of \sysname. The fine-grained search (FGS Ref. PC) and coarse-grained search (CGS Ref. PC) against the \tcloud are the most time-consuming stages, taking about 360 ms and 250 ms on average, respectively. Together, they account for roughly 69\% of the total latency. Although the fine-grained search processes fewer points, it requires more time because it performs a detailed KD-tree lookup, whereas the coarse-grained search operates at a lower resolution. 
% \textcolor{blue}{
The fine-grained search with the map (FGS Map) is faster than the FGS Ref. PC because it matches a relatively smaller set of points. 
% }
Draco and LZMA compression contribute only a small fraction of the total latency.
% \textcolor{blue}{
% The KD-tree structure that \sysname builds for the 3D map for fast s
\sysname builds the KD-tree structure from the 3D map offline. 
On average, for a 3D map of 1.5~GB, \sysname takes 17~seconds to build the KD-tree structure.
% }

\figref{fig:throughput} shows the throughput of \sysname and three baselines as a function of the compression ratio. Throughput is defined as the number of point clouds compressed and transferred to cloud storage per second.
For GPCC, Draco, and Octree, the throughput remains relatively constant across compression ratios. In contrast, \sysname’s throughput increases with higher compression ratios. This is because larger \textit{distance thresholds} in \sysname identify more common points, reducing the number of exclusive points that require fine-grained searches. As a result, the compression latency decreases, leading to higher throughput.
Despite this improvement, \sysname’s throughput is still approximately 2.2× lower than Draco’s, as it trades off latency to achieve better compression ratios.
% \begin{figure*}[ht]
%     \centering
%     % Subfigure 1
%     \begin{subfigure}{0.32\textwidth}
%         \centering
%         \includegraphics[width=\textwidth]{figures/compression_fps.pdf}
%         \caption{Compression FPS}
%         \label{fig:compression_fps}
%     \end{subfigure}
%     % Subfigure 2
%     \begin{subfigure}{0.32\textwidth}
%         \centering
%         \includegraphics[width=\textwidth]{figures/decompression_fps.pdf}
%         \caption{Decompression FPS}
%         \label{fig:decompression_fps}
%     \end{subfigure}
%     % Subfigure 3
%     \begin{subfigure}{0.32\textwidth}
%         \centering
%         \includegraphics[width=\textwidth]{figures/e2e_fps.pdf}
%         \caption{End-to-end FPS}
%         \label{fig:end-to-end_FPS}
%     \end{subfigure}
%     % % Subfigure 4
%     % \begin{subfigure}{0.24\textwidth}
%     %     \centering
%     %     \includegraphics[width=\textwidth]{figures/dynamic_env_1.pdf}
%     %     \caption{Dynamic Content}
%     %     \label{fig:dynamic_content}
%     % \end{subfigure}
%     \caption{Learning-based methods are not able to maintain higher FPS at higher compression ratios, which is not desirable for autonomous vehicle applications}
%     \label{fig:sensitivity_analysis}
% \end{figure*}
% \vspace{-}

\subsection{Comparison with Learning-based Methods}
\label{ss::learning_based_methods}

\begin{figure}[t]
    \vspace{-15pt}
    \centering
    \includegraphics[width=0.7\columnwidth]{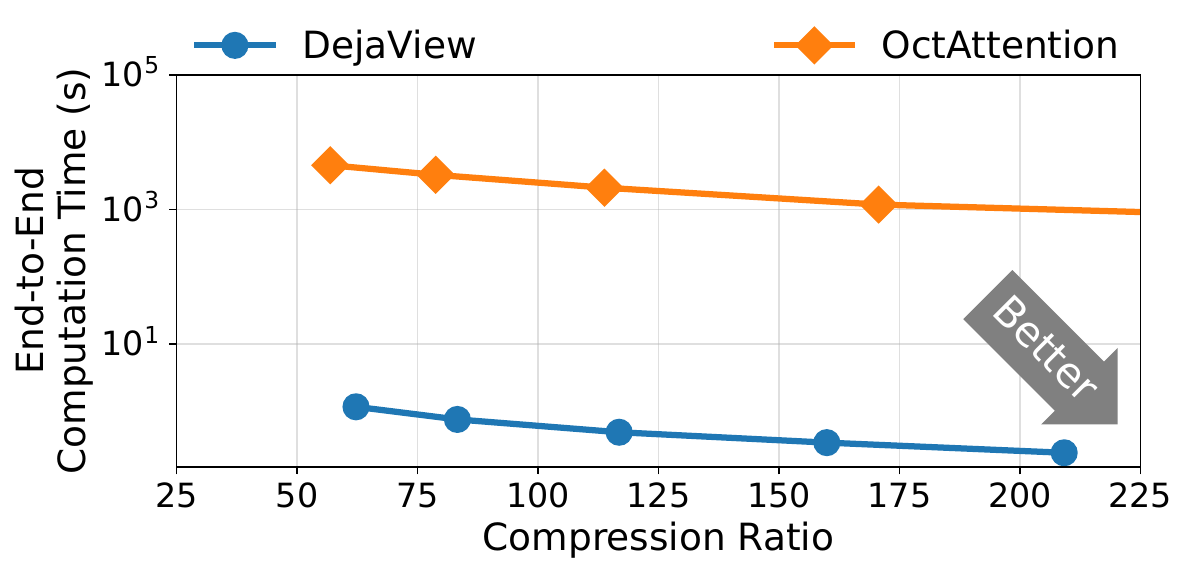} 
    \caption{The y-axis is in logarithmic scale. A learning-based compression model (OctAttention) requires three orders of magnitude more time to compress and decompress point clouds relative to \sysname. }
    \label{fig:e2e_time}
\end{figure}

Recent learning-based point cloud compression techniques, such as OctAttention~\cite{fu2022octattention} and OctSqueeze~\cite{huang2020octsqueeze}, leverage Octree representations and context models for entropy encoding.
These methods can achieve better reconstruction quality at a given compression ratio, but the time required to compress and decompress point clouds is very high~\cite{you2025reno}. 
To evaluate this tradeoff, using over $2000$ point clouds, we measured the compression ratio, reconstruction accuracy (Chamfer distance), and end-to-end computation time for \sysname and OctAttention. 
End-to-end computation time is the sum of compression and decompression latency.
We used the authors' open source OctAttention model~\cite{git-octattention} pre-trained on the KITTI dataset~\cite{behley2019semantickitti}. 
To simulate a realistic deployment scenario, we attached a P1000 GPU to our edge computing platform and ran inference on that.
% running inference on a P1000 GPU that we attached to our edge computing platform.
% To simulate a practical deployment scenario, we attached a P1000 GPU to our edge computing platform.
 % running on a single GPU. This likely underestimates its best-case performance, but reflects practical deployment latency in AV systems where fast, lightweight compression is preferred.
We plot the results for this experiment in \figref{fig:e2e_time}. 
The x-axis represents the compression ratio. 
The y-axis, on a logarithmic scale, represents end-to-end computation time, which is the sum of compression and decompression times.

Across all compression ratios, on average, OctAttention achieves a higher reconstruction accuracy than \sysname \ie 5~cm lower Chamfer distance.  
For higher compression ratios \ie 200, the reconstruction accuracy difference is only 1~cm.
However, this comes at the cost of end-to-end computation time.
\textit{For a given compression ratio, the end-to-end computation time for OctAttention is approximately 3990~times more than \sysname (\figref{fig:e2e_time})!}
Putting this number in perspective, for a compression ratio of 170, it takes \sysname 340~ms to compression and decompress a single point cloud.
% compression and decompression times sum up to 340~ms for a single point cloud. 
For OctAttention, on the other hand, it takes almost 9~minutes to compress and decompress the same point cloud.
% the sum of compression and decompression time is almost 9~minutes for a single point cloud.
\textit{This means that for one second of data from a single vehicle-mounted LiDAR, it would take OctAttention 1.5~hours to compress and decompress the 3D data.} 
% Consequently, unlike \sysname, learning-based point cloud compression techniques such as OctAttention are less suitable for the application use cases mentioned in this paper.
% Although learning-based approaches such as OctAttention are promising, their high computational cost limits their practicality for the application use cases mentioned in this paper.
Although deep learning techniques achieve comparable compression ratios with marginally better reconstruction accuracy, their three orders of magnitude higher latency renders them impractical for the time-sensitive use cases that we target. 
In contrast, \sysname trades a small amount of reconstruction accuracy for faster computation times and comparable compression ratios.
% modestly for large compression gains and faster computation times.

%% file: 5_related.tex
\section{Related Work}
\label{s:rel}

\parab{Traditional Methods. }Traditional compression algorithms leverage spatial redundancies in point clouds using hierarchical tree structures. Among these, Octree is the widely used~\cite{octree, schwarz2018emerging, introduction_to_standards}. Octree-based methods quantize the point cloud and then recursively subdivide it into eight cubes and encode their occupancy for compression.
In contrast, \sysname employs an Octree structure for coarse-grained \textit{diff} operation but mitigates the artifacts caused by quantization and binning through a fine-grained \textit{diff} operation that refines the results at a point level. 
Furthermore, Octree representations have also been utilized to capture spatial redundancies between consecutive point clouds to improve compression ratios~\cite{octree2buff, garcia2018intra, de2016compression, lidar_compression_survey, dricot2019hybrid}. 
\sysname, on the other hand, identifies spatial redundancies across larger temporal scales by finding a \tcloud in the vicinity of \scloud, to achieve a higher compression ratio.
% }

% \textcolor{blue}{
While Octree-based methods are effective for sparse point clouds, KD-tree-based approaches are often preferred for denser point clouds. Unlike the octree, the KD-tree is primarily used to reorder points so that spatially adjacent points are stored close together in memory. This spatial reordering improves the compression ratio by enhancing the performance of entropy and predictive coding techniques~\cite{kdtree_devillers2000geometric}. The open-source compression library Draco from Google~\cite{Draco} leverages the KD-tree for this purpose. \sysname, on the other hand, uses KD-tree for efficiently computing fine-grained \textit{diff} between point clouds.
\parab{Learning-based Methods. }Early learning-based methods voxelize point clouds and use convolutional autoencoders to encode the 3D grids~\cite{quach2019learning, wang2021lossy, wang2021multiscale}. However, voxelization is computationally expensive and memory-intensive, as it allocates voxels even for empty regions. To address this limitation, subsequent works such as PointNet and PointNet++ directly process raw point clouds without voxelization~\cite{qi2017pointnet, qi2017pointnet++, huang20193d, yan2019deep}. Another line of research projects point clouds into 2D or range images, which are then compressed using image-based techniques~\cite{ainala2016improved, he2020point, houshiar20153d, tu2016compressing, tu2019point, feng2020real, zhou2022riddle}. While these methods effectively exploit local spatial and temporal correlations, they primarily capture short-term redundancies between consecutive frames. In contrast, \sysname identifies and leverages long-term spatial and temporal redundancies, achieving higher compression ratios without the heavy computational cost of voxelization or network training.
% }

% \textcolor{blue}{
Hybrid methods combine deep learning with conventional hierarchical structures such as octrees. These approaches enhance compression by learning entropy models that exploit the spatial and hierarchical context of octree nodes, including their level, occupancy, and neighborhood relationships~\cite{huang2020octsqueeze, que2021voxelcontext, fu2022octattention, cui2023octformer, song2023efficient}. While such methods achieve high compression ratios, they incur significant computational overhead and require training separate models for each dataset, limiting their real-time applicability and generalization~\cite{you2025reno}. In contrast, \sysname leverages hierarchical representations to efficiently compute differences between point clouds without the overhead of model training, while maintaining generalizability
% }

% \ali{shepherd+reviewer:  Clarify how the proposed method differs from the well-established practice of LiDAR-based SLAM}
% \fawad{Addressed.}

% \textcolor{blue}{
\parab{LiDAR-based SLAM. }SLAM processes a stream of point clouds to estimate a vehicle’s trajectory and maintains a single consistent 3D map of the environment~\cite{fastlio, fast-lio2}. 
To achieve this, SLAM continuously performs point cloud-to-map alignment to refine vehicle pose. 
In doing so, it also computes residual differences between the current point cloud and the existing map.
% As it does this, SLAM performs point cloud to map alignment to localize a vehicle and compute the difference  a \textit{diff} operation to identify the differences between each point cloud and the 3D map. 
Of these differences, it integrates static and stable 3D points into the map. 
SLAM discards unstable and dynamic points (such as those belonging to vehicles, pedestrians \etc). 
Consequently, it does not preserve raw point clouds and cannot reconstruct them from the processed 3D map.
% }

% \textcolor{blue}{
% In contrast, \sysname operates offline with known vehicle poses to compress raw point clouds. 
Unlike SLAM, \sysname does not distinguish between static and dynamic points. 
Instead, it performs a bidirectional \textit{diff} operation to identify common and exclusive points. 
It stores references to common points and compactly encodes exclusive points. 
% Unlike SLAM which merges and discards point clouds, 
\sysname preserves every point cloud, including dynamic content, in a fully reconstructible form.
This enables downstream applications such as forensic replay, safety validation, and retraining of perception systems, and can even provide data to build SLAM maps.
% Conceptually, SLAM’s \textit{diff} is pose-driven and used for state estimation, whereas our \textit{diff} is data-driven and used for redundancy elimination and reconstruction fidelity.
% }

%% file: 7_future_work.tex
% \vspace{-4pt}
\section{Discussion and Future Work}
% \ali{ reviewer:
% When is the reference dataset updated? Is it updated weekly? What are the selection rules for determining whether a point cloud should be included? (Rev E)
% The paper seems to assume that the 3D map and the reference dataset remain static. However, in real-world scenarios, road maintenance, demolition of old buildings, and the construction of new ones require constant updates to the AV's 3D maps. (Rev A) 
%  It would be helpful to include optimization experiments on how many reference clouds to keep and at what spatial or pose interval to keep them.
% }
% \textcolor{blue}{
Existing approaches exploit redundancies across short temporal intervals (i.e., between consecutive frames) to compress point clouds. \sysname, on the other hand, leverages a 3D map and spatially proximate point clouds from a reference dataset to identify redundancies over much larger temporal scales (days, months, or even years), achieving significantly higher compression ratios.
\sysname performs best when the source data closely matches the reference dataset and 3D map. However, changes such as road maintenance, construction, or building demolition can render the reference dataset and 3D map stale, thereby reducing the compression ratio. Regularly updating the reference dataset and 3D map can help maintain high compression performance. 
% Though it is straightforward to compres
For previously compressed data, \sysname can adopt one of two strategies: it can either recompress the data with respect to the updated reference dataset and 3D map, or maintain separate reference datasets and maps for different sets of source data. We leave an in-depth examination of these and other innovative approaches to future work.
% Developing a systematic approach to detect environmental changes, update references, and adjust past compressed data to preserve its validity remains an open direction for future work.
% }

% \textcolor{blue}{
The selection of point clouds that make up the reference dataset provides another avenue for improvement for \sysname. 
For the purposes of this paper, we used the first 10 days of collected data to build the reference dataset. 
Future work can explore intelligent point cloud selection techniques, across space and time, that optimize compression ratio and reduce, if not eliminate, the need to update the reference dataset altogether.
% Moreover, and reduce, it not eliminate, the need to update the reference dataset.
% update times
% In this paper, we built the the reference dataset using the first 10 days of collected data. Future work can explore techniques to intelligently select point clouds across space and time for the reference dataset. 
% In future work, we plan to investigate strategies for dynamically updating the reference dataset and 3D map, as well as optimizing reference data selection to improve both compression efficiency and storage management.
% }

%% file: 6_conclusion.tex
\section{Conclusion}

% LiDAR is an important sensor in AVs that provides accurate 3D information about the environment. A major portion of the data produced by AVs comes from LiDAR. This data needs to be transferred over the network to cloud storage. The huge volume of uncompressed LiDAR data places a substantial burden on network bandwidth and increases storage costs. Existing compression techniques address inter-frame redundancies but do not fully exploit the redundancies present over longer time frames. As AVs operate within limited areas and frequently pass through the same regions, redundancies also accumulate over extended periods, such as days, weeks, or even months. To address this, we propose a novel system called \sysname that accounts for redundancies over these longer time frames, achieving better compression ratios and reconstruction quality compared to existing methods.

In this paper, we introduce \sysname a novel methodology to compress AV point clouds leveraging multiple traversals. Through experimentation, we compared our proposed compression algorithm with state-of-the-art methods on two datasets. One was collected in the real world using an Ouster OS1-128 beam LiDAR, and the other was logged from a photo-realistic autonomous driving simulator, CARLA. In both datasets, we achieve 2.5x more compression compared to the SOTA method. We also conducted experiments for localization, 3D object detection, and 3D semantic segmentation to demonstrate the effectiveness and validity of \sysname on downstream tasks in the autonomous driving pipeline.

%% file: 8_Acknowledgment.tex
\parab{Acknowledgment.} 
We thank the anonymous reviewers and shepherd for their insightful comments.
This material is based upon work supported
by the U.S. National Science Foundation under grants CNS-2348461.